\documentclass[a4paper,11pt]{article}

\usepackage{jcappub} 

\usepackage{placeins}
\usepackage[utf8]{inputenc}
\usepackage{graphicx}
\usepackage{tensor}
\usepackage[normalem]{ulem}
\usepackage{verbatim}
\usepackage{epsfig}
\usepackage{subfigure}
\usepackage{color}
\usepackage{multirow}
\usepackage{comment}
\usepackage{slashed}
\usepackage{amsmath}
\usepackage[normalem]{ulem}
\usepackage{framed}

\usepackage{amssymb,amsmath,amsfonts,mathrsfs}
\usepackage[dvipsnames]{xcolor}
\usepackage{graphicx}
\usepackage{longtable}
\usepackage{verbatim}
\usepackage{color}
\usepackage[mathscr]{euscript}
\usepackage{framed}
\usepackage{mdframed} 

\usepackage{cancel}
\usepackage{color}
\usepackage{hyperref}
\hypersetup{colorlinks, citecolor=bluscuro, linkcolor=black, urlcolor=bluscuro}
\definecolor{rossos}{cmyk}{0,1,1,0.55}
\definecolor{bluscuro}{rgb}{0.15, 0.2, .85}
\definecolor{bluchiaro}{cmyk}{1,.3,0.,0.1}

\numberwithin{equation}{section}

\newcommand{\be}{\begin{equation}\begin{aligned}}
\newcommand{\ee}{\end{aligned}\end{equation}}
\def\ii{{\text{\tiny i}}}
\def\d{{\rm d}}
\def\co{{\text{\tiny cut-off}}}

\newcommand{\Msun}{\ensuremath{M_\odot}}

\newcommand{\beq}{\begin{equation}}
\newcommand{\eeq}{\end{equation}}
\newcommand{\bit}{\begin{itemize}}
\newcommand{\eit}{\end{itemize}}

\newcommand{\ddat}{\boldsymbol{\mathcal{D}}} 
\newcommand{\pp}{\mathcal{P}} 
\newcommand{\tth}{\boldsymbol{\vartheta}} 
\newcommand{\mdl}{\mathcal{M}}

\newcommand{\nn}{\boldsymbol{\mathcal{N}}}

\usepackage{tikz}

\title{The Importance of Priors on LIGO-Virgo Parameter Estimation: the Case of Primordial Black Holes}

\author[a]{S.~Bhagwat,}
\author[b]{V.~De~Luca,}
\author[b]{G.~Franciolini,}
\author[a,c]{P.~Pani,}
\author[b,c]{A.~Riotto}

\affiliation[a]{Dipartimento di Fisica, “Sapienza” Università di Roma, Piazzale Aldo Moro 5, 00185, Roma, Italy}
\affiliation[b]{
	Department of Theoretical Physics and Center for Astroparticle Physics (CAP) \\
			24 quai E. Ansermet, CH-1211 Geneva 4, Switzerland}
\affiliation[c]{INFN, Sezione di Roma, Piazzale Aldo Moro 2, 00185, Roma, Italy}

\abstract{The black holes detected by current and future interferometers can have diverse origins. Their expected mass and spin distributions depend on the specifics of the formation mechanisms.  
When a physically motivated prior distribution is used in a Bayesian inference, the parameters estimated from the gravitational-wave data can change significantly, potentially affecting the physical interpretation of certain gravitational-wave events and their implications on theoretical models.
As a case study we analyze primordial black holes, which might be formed in the early universe and could comprise at least a fraction of the dark matter. If accretion is not efficient during their cosmic history,  primordial black holes are expected to be almost non-spinning. If accretion is efficient, massive binaries tend to be symmetrical and highly spinning.
We show that incorporating these priors can significantly change the inferred mass ratio and effective spin of some binary black hole events, especially those identified as high-mass, asymmetrical, or spinning by a standard analysis using agnostic priors. 
For several events, the Bayes factors are only mildly affected by the new priors, implying that it is hard to distinguish whether merger events detected so far are of primordial or astrophysical origin.
In particular, if binaries identified by LIGO/Virgo as strongly asymmetrical (including GW190412) are of primordial origin, their mass ratio inferred from the data can be closer to unity. For GW190412, the latter property is strongly affected by the inclusion of higher harmonics in the waveform model.
}

\emailAdd{swetha.bhagwat@roma1.infn.it}
\emailAdd{valerio.deluca@unige.ch}
\emailAdd{gabriele.franciolini@unige.ch}
\emailAdd{paolo.pani@uniroma1.it}
\emailAdd{antonio.riotto@unige.ch}

\begin{document}
\maketitle
\flushbottom

\section{Introduction}
\label{sec:intro}
Following the historical breakthrough by the LIGO/Virgo Collaboration (LVC), gravitational-wave (GW) astronomy is now in full blossom.
As of October 2020, the LIGO-Virgo interferometers have so far provided us with at least 13 GW observations of candidate black-hole~(BH) binaries~\cite{LIGOScientific:2018mvr,LIGOScientific:2020stg}. Further events from the first part of the third observational run~(O3a) have been released very recently~\cite{Abbott:2020niy}, and many more are expected in the future.
Given the current and upcoming wealth of data, it becomes of utmost importance to confront  specifics of formation models with observations and to infer the (astro)physical implications. As the number of detections increases, we will soon be in a position to address important questions like whether the observed binary BH population has the same origin or multiple formation channels.

One intriguing hypothesis is that (at least a fraction of) the merging BHs are of primordial origin, i.e., they formed early in the evolution of the universe (see Refs.~\cite{Sasaki:2018dmp,Green:2020jor} for recent reviews). This possibility is also interesting as primordial BHs~(PBHs) may comprise the totality or a fraction of the Dark Matter~(DM) in the universe~\cite{Bird:2016dcv,Sasaki:2016jop,Carr:2020gox}.

Exploring the primordial origin of GW events has motivated several works on the PBH formation mechanisms~\cite{Blinnikov:2016bxu,Ivanov:1994pa,GarciaBellido:1996qt,Ivanov:1997ia}, 
merger rates~\cite{Bird:2016dcv,Sasaki:2016jop,Ali-Haimoud:2017rtz,Raidal:2018bbj,Hutsi:2019hlw,
Vaskonen:2019jpv,Jedamzik:2020ypm,Jedamzik:2020omx,Young:2020scc,Clesse:2020ghq}, confrontation with other astrophysical constraints~\cite{Allsman:2000kg, Kashlinsky:2016sdv, Ali-Haimoud:2016mbv, Oguri:2017ock,Carr:2017jsz,bellomo,Zumalacarregui:2017qqd,
Manshanden:2018tze,
Niikura:2019kqi,
Murgia:2019duy,
Chen:2019xse,
Serpico:2020ehh,
DeLuca:2020fpg,
Lu:2020bmd}, and with current GW data~\cite{Clesse:2016vqa,Chen:2018czv,Raidal:2018bbj,Gow:2019pok,DeLuca:2020bjf, Dolgov:2020xzo,DeLuca:2020qqa,Clesse:2020ghq}.
Nonetheless, all previous studies have adopted the measurements (in particular of the masses and spins of the binary BH components and source redshift) obtained by the Bayesian inference that uses conventional agnostic prior distributions for the binary parameters (see Sec.~\ref{sec:analysis} below for details).
It is well-known that a drastically different choice of priors can impact the inferred value of the parameters (i.e., their posterior distributions). In some cases, different prior assumptions might significantly change the physical interpretation of GW events~\cite{Vitale:2017cfs,Mandel:2020lhv,Zevin:2020gxf}.
Thus, when the parameters of the observed BHs are used to constrain, calibrate, and test physical models, the effects of priors on the inference of the parameters need to be quantified. 

In this paper, we investigate the impact of the priors under the hypothesis that (at least some of) the GW events detected so far are of primordial origin. We  argue that the first crucial ingredient to explore any consequence of current and future GW events for the PBH scenario is to make sure that the values of the parameters inferred from GW data (in particular BH masses and spins) are not affected by a PBH-motivated (as opposed to an agnostic) prior.  As we shall show, this is indeed the case for several GW events, but not for all.  We find that a PBH-motivated prior on the masses and spins can significantly affect the parameter estimation, especially the measurement of the binary mass ratio and effective spin, for those BH binaries identified as being either high-mass, or asymmetrical, or spinning using the agnostic priors.
The underlying reason for this is twofold: i) the initial mass distribution of PBHs is not uniform across the mass range but depends on the physical formation mechanism that occurs in the early universe; ii) more importantly, the natal spin of PBHs is negligible, essentially restricting some of the parameters (most notably the effective spin $\chi_\text{\tiny eff}$) in the waveform model. Unless accretion is efficient in spinning up the PBHs during their history (see Sec.~\ref{sec:priors}), a negligible-spin prior strongly affects the recovered parameters for those binaries identified as spinning by the standard LVC analysis. In particular, the inference on the mass ratio of the binary BHs are significantly modified to compensate for the absence of $\chi_\text{\tiny eff}$ in the waveform and due to the different prior distribution of the masses.
Furthermore, if the PBH accretion is efficient, the binaries made of heavy PBHs tend to be symmetrical and highly-spinning. This affects the posterior distributions for those binary BH events which are identified as asymmetrical or non-spinning with agnostic priors.
Overall these effects can have deep consequences for the physical interpretation of some events, especially for systems like GW190412, which is identified as an asymmetrical, spinning binary using standard priors~\cite{LIGOScientific:2020stg}.

We note that while we focus on the PBH hypothesis as a case study, our results for the non-accreting case are more general and indeed apply to other scenarios (possibly of astrophysical~\cite{Fuller:2019sxi} or quantum~\cite{Bianchi:2018ula} origin) in which the spin of the binary components is expected to be small.
 
The rest of this paper is organized as follows. In Sec.~\ref{sec:priors} we review the PBH mass and spin distributions that should be consistently adopted if one wishes to investigate the PBH scenario. Section~\ref{sec:analysis} reviews aspects of the Bayesian parameter estimation and explains our method. Our results are presented and discussed in Sec.~\ref{sec:results}, whereas Sec.~\ref{sec:conclusion} is devoted to our conclusions. 

\section{PBH mass and spin distributions}
\label{sec:priors}
In this section we present a short summary of the main theoretical predictions for the mass and the spin distributions of PBHs. We will first focus on the scenario where no accretion is taking place, and then we analyse the consequences of a phase of accretion in the PBH cosmological evolution. The reader can find additional details in Refs.~\cite{DeLuca:2020bjf,DeLuca:2020qqa}.

\subsection{Without accretion}
The formation of a significant population of PBHs in the universe has been addressed by several models in the literature \cite{Sasaki:2018dmp}. One of the most promising possibility arises in scenarios where PBHs are formed from the collapse of sizeable overdensities in the radiation dominated era~\cite{Blinnikov:2016bxu,Ivanov:1994pa,GarciaBellido:1996qt,Ivanov:1997ia}. The PBH mass distribution at the formation epoch $z_\ii$ is determined by the intrinsic properties of the collapsing density perturbations, dictated by the inflationary curvature perturbations, and is usually parametrized by a lognormal distribution,
\begin{equation}
\label{psi}
\psi (M,z_\ii) = \frac{1}{\sqrt{2 \pi} \sigma M} {\rm exp} \left(-\frac{{\rm log}^2(M/M_c)}{2 \sigma^2} 
\right)\,,
\end{equation}
in terms of its width $\sigma$ and central mass scale $M_c$. This shape is often introduced to describe several models and originates when the power spectrum of the curvature perturbations generated during inflation has a symmetric peak at the small scales giving rise to the large overdensities needed to generate the PBHs upon horizon re-entry~\cite{Dolgov:1992pu,Carr:2017jsz}. 
We use the label "i" to indicate quantities at formation epoch.

In this common formation scenario, the nearly-spherical shape~\cite{bbks} of the collapsing perturbations and the small collapsing time in the radiation dominated era implies that the PBH initial spin $\chi_\ii \equiv |J_\ii|/M_\ii^2$ should be below the percent level~\cite{DeLuca:2019buf,Mirbabayi:2019uph}
\be
\label{a}
\chi_{\text{\tiny i}} \sim 10^{-2} 
\sqrt{1-\gamma^2},
\ee
expressed as a function of the effective width of the power spectrum $\gamma$~\cite{DeLuca:2019buf}. For example, a very peaked curvature perturbation power spectrum leads to a very narrow PBH mass function and to   $\gamma\simeq 1$, thus  further suppressing the initial PBH spin.   

In the scenario in which PBHs do not undergo a phase of accretion in their cosmological evolution, their initial mass and spin distributions will be preserved up to the present epoch, as given by Eqs.~\eqref{psi} and \eqref{a}, respectively. They therefore determine  the properties of the population at the time of merger.

\subsection{With accretion} 
\label{sec:priors_accr}
Throughout the cosmological history, there may be periods in which baryonic mass accretion affects PBHs in binary systems, impacting their mass function~\cite{Ricotti:2007jk,Ricotti:2007au,zhang} and spins~\cite{bv,DeLuca:2020bjf,DeLuca:2020qqa}. This happens for binaries with mass $M_\text{\tiny tot}\gtrsim\mathcal{O}(10) M_\odot$, for which the binary as a whole attracts the surrounding gas as long as its typical size is smaller than its corresponding Bondi radius~\cite{DeLuca:2020bjf, DeLuca:2020qqa}. In this situation, both the PBHs accrete baryonic particles, with individual rates determined by their orbital velocities and their masses.
The individual accretion rates are given in terms of the  binary mass ratio $q \equiv M_2/M_1 \leq 1$ as
\begin{align}
\dot M_1 = \dot M_\text{\tiny bin}  \frac{1}{\sqrt{2 (1+q)}}, \qquad \dot M_2 = \dot M_\text{\tiny bin}  \frac{\sqrt{q} }{\sqrt{2 (1+q)}},\label{M1M2dotFIN}
\end{align}
as a function of the Bondi-Hoyle mass accretion rate for the binary system
\be \label{R1bin}
\dot M_\text{\tiny bin} = 4 \pi \lambda m_H n_\text{\tiny gas} v^{-3}_\text{\tiny eff} M^2_\text{\tiny tot}.
\ee
The latter is driven by the total mass $M_\text{\tiny tot} = M_1 + M_2$ and depends on the binary effective velocity $v_\text{\tiny eff}$ relative to the baryons with cosmic mean density $n_\text{\tiny gas}$ and the hydrogen mass $m_H$.
The effects of the Hubble expansion, the gas viscosity, and the coupling of the CMB radiation to the gas through Compton scattering are tracked by the accretion parameter $\lambda$, whose explicit expression can be found in Ref.~\cite{Ricotti:2007jk}. 
The observational constraints on the fraction of PBHs as DM, for masses larger than $\mathcal{O}(M_\odot)$, imply that a secondary DM-component halo should form around the isolated or binary PBHs~\cite{Ricotti:2007au,Adamek:2019gns,Mack:2006gz}. Its main effect is to catalyze the gas accretion rate and is usually tracked into the accretion parameter $\lambda$ (details are provided in App.~B of Ref.~\cite{DeLuca:2020bjf} and references therein).

We draw the reader's attention to the fact that the accretion phase depends on the complex phenomena related to the local, global~\cite{Ricotti:2007au,Ali-Haimoud:2016mbv} and mechanical~\cite{Bosch-Ramon:2020pcz} feedbacks, X-ray pre-heating~\cite{Oh:2003pm} and the formation of structures~\cite{Hasinger:2020ptw,raidalsm,Ali-Haimoud:2017rtz}. Overall, these effects can make the accretion much less efficient at a relatively small redshift. We have decided to parametrize these uncertainties into a cut-off redshift $z_\text{\tiny cut-off}$ below which accretion is negligible~\cite{DeLuca:2020bjf,DeLuca:2020qqa}, with representative values $z_\text{\tiny cut-off}= \{ 15,10,7 \}$. A smaller cut-off redshift is associated to a prolonged accretion phase. In the analysis of this section we include $z_\text{\tiny cut-off}=7$ for completeness, although such a value corresponds to a very strong accretion phase that is disfavored~\cite{DeLuca:2020bjf,DeLuca:2020qqa}.

The PBH evolution is strongly affected by accretion in several aspects: 
\begin{enumerate}
\item it induces a change of the PBH mass function according to the equation
\begin{equation}
\psi(M(M_\ii,z),z) \d M = \psi(M_\ii,z_\ii) \d M_\ii,
 \label{psiev}
\end{equation}
where $M(M_\ii,z)$ is the final mass at redshift $z$ for a PBH with mass $M_\ii$ at redshift $z_\ii$.
The evolved mass distribution is broader at higher masses and has a high-mass tail that is orders of magnitude larger than its corresponding value at formation \cite{DeLuca:2020bjf};

\item it modifies the fraction of PBHs in the DM depending on the redshift with important consequences when confronting the existing constraints with the physical parameters, see Ref.~\cite{DeLuca:2020fpg} for details;

\item it pushes the mass ratio towards unity as $\dot q/q = ( \dot M_2/M_2 -  \dot M_1/M_1 ) > 0$,
since the secondary binary component always inherits a larger accretion; 

\item it influences the PBH spin evolution. Indeed, the geometry of the accretion process and the PBH spins may be crucially impacted by the angular momentum that the infalling gas particles carry~\cite{bv}. In particular, the formation of a thin accretion disk~\cite{Ricotti:2007au,Shakura:1972te, NovikovThorne} leads to an efficient angular momentum transfer, with the consequent growth of the PBH spins with mass accretion as 
\begin{equation}
\dot \chi = g(\chi) \frac{\dot {M}}{M},
\end{equation}
in terms of a function  $g(\chi)$ of the dimensionless Kerr parameter (see Refs.~\cite{Bardeen:1972fi,Brito:2014wla,volo,DeLuca:2020qqa}), until it reaches the limit $\chi_\text{\tiny max} = 0.998$ dictated by radiation effects~\cite{thorne,Gammie:2003qi}. The larger accretion rates inherited by the secondary component of the binary results into a larger growth of its spin with respect to the one of the primary component.
\end{enumerate}

\subsection{Physically-motivated distributions of the PBH binary parameters} \label{sec2.3}

The conventional way of inferring the physical parameters of a GW model and their uncertainties is using a Bayesian inference setup, which relies on combining prior expectations of the parameter values with the likelihood  from the observation to obtain a posterior distribution for the parameters (see Sec.~\ref{sec:analysis} for details). 
In this section, we discuss the prior distributions on the relevant binary BH parameters under the hypothesis that GW events are of primordial origin.

\begin{figure*}[t!]
\includegraphics[width=0.9\textwidth]{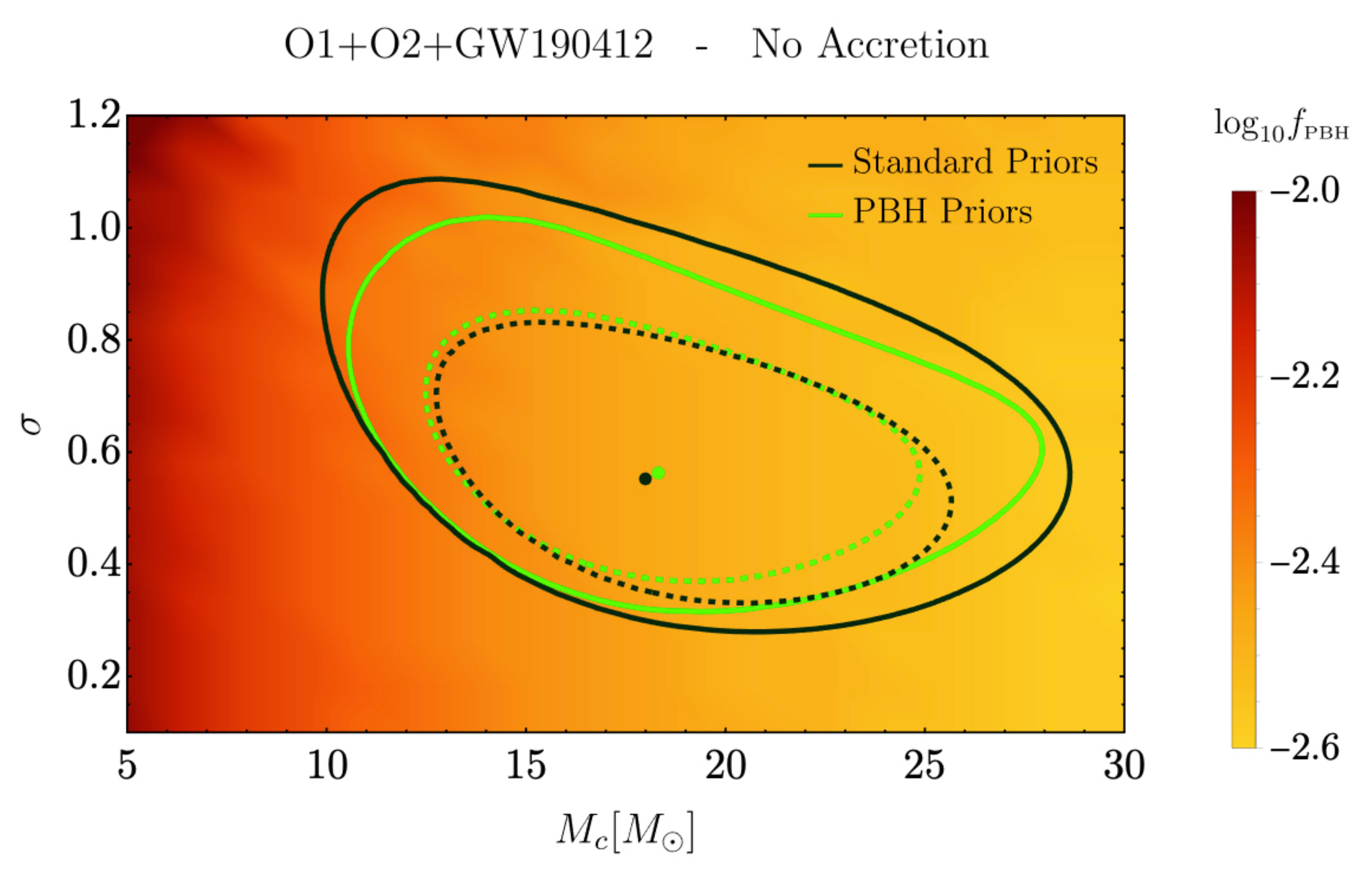}
 \caption{$\chi^2$-analysis using binary parameters (masses and source redshift) obtained with both the standard (flat) priors used by the LVC analysis and the PBH-motivated priors discussed in the following (for the case of no accretion, see Table~\ref{tab:Log Bayes Factor}). Dashed (solid) contours represent the $2\sigma$ ($3\sigma$) confidence interval. The estimation of $M_c$ and $\sigma$ is only mildly affected by the different choices of the priors. Here $f_\text{\tiny PBH}$ is the fraction of PBHs in DM, see Ref.~\cite{DeLuca:2020qqa} for details. }  
 \label{fig:likelihoodMcs}
 \end{figure*}

Due to the lack of a preferred specific set of parameters for the PBH mass function, we  make the assumption that all the BH coalescences detected so far are of primordial origin\footnote{By analyzing the distribution of the masses and redshift, a recent analysis of O1 and O2 events argues that this case is disfavored compared to the astrophysical formation scenario~\cite{Hall:2020daa}. It would be interesting to quantify whether the inclusion of accretion and/or other parameters such as
the spins in the analysis can change this result.} and determine the phenomenological values of $M_c$ and $\sigma$ in Eq.~\eqref{psi} which best-fit the data by performing a $\chi^2$-analysis~\cite{DeLuca:2020qqa}. Such values can be taken as the starting point of our study.
When performing the analysis to find the best model for the distribution of PBH mass and spin, we used the LVC data, which adopted non-informative priors (see Sec.~\ref{sec:analysis} for details). Anticipating some results discussed later on, we find that some of the posterior distributions are affected by the choice of PBH-motivated priors, which in turn can potentially modify the best-fit mass function. However, as shown in Fig.~\ref{fig:likelihoodMcs}, we have checked that the best-fit values are stable with respect to the changes induced by re-analyzing each individual events with the PBH-motivated priors discussed in the rest of this work. This shows that our procedure is consistent. The resulting observable distributions described in Ref.~\cite{DeLuca:2020qqa} are not affected by the PBH-motivated choice of priors. Notice also that, since the selected values of 
$f_\text{\tiny PBH}$ are below $10^{-2}$, the role of PBH clustering can be neglected \cite{veermae}.

After fixing the PBH model parameters with the values calibrated using the data, we  now discuss the expected distributions of the physical GW observables. The easiest parameter that can be extracted from a binary BH coalescence waveform is its chirp mass,
\begin{equation}
M_\text{\tiny chirp} = \frac{(M_1 M_2)^{3/5}}{(M_1 + M_2)^{1/5}},
\end{equation}
which affects the leading-order Newtonian GW phase~\cite{Blanchet:2013haa}. The first post-Newtonian correction includes the mass ratio $q$, whereas spin-angular momentum couplings enter to next-to-the-leading order, mainly through
the binary's effective spin parameter,
\begin{equation}
\label{chieff}
\chi_\text{\tiny eff} \equiv \frac{\chi_1 \cos{\theta_1} + q \chi_2 \cos{\theta_2}}{1+q}\,,
\end{equation}
in terms of the individual BH spins $\chi_j$ ($j=1,2$), and the angles $\theta_1$ and $\theta_2$ between the orbital angular momentum and the individual spin vectors. In the PBH scenario, we expect the spin vectors to be uniformly distributed on the two-sphere.

\begin{figure*}
 \includegraphics[width=0.49\textwidth]{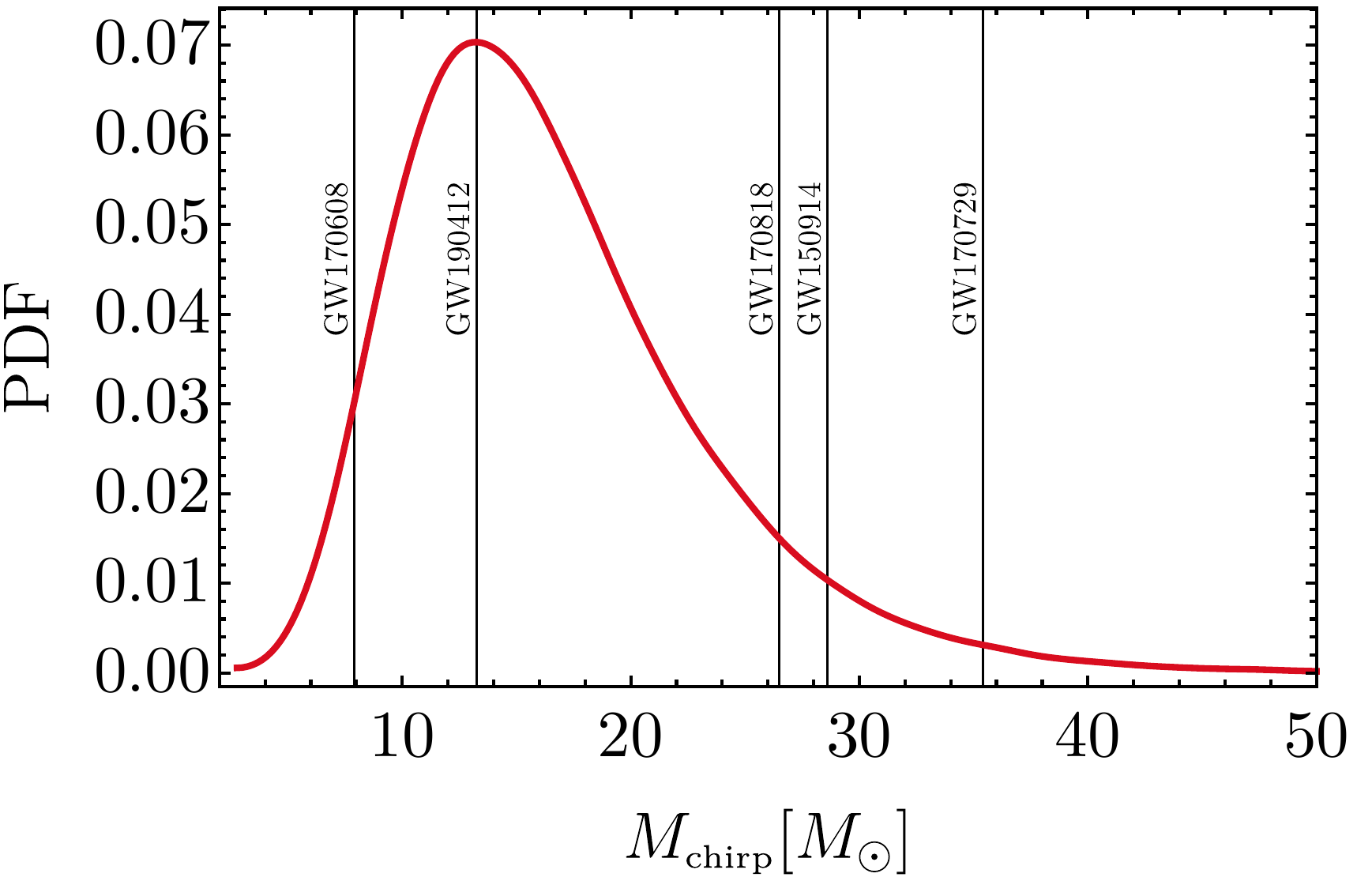}
  \includegraphics[width=0.49\textwidth]{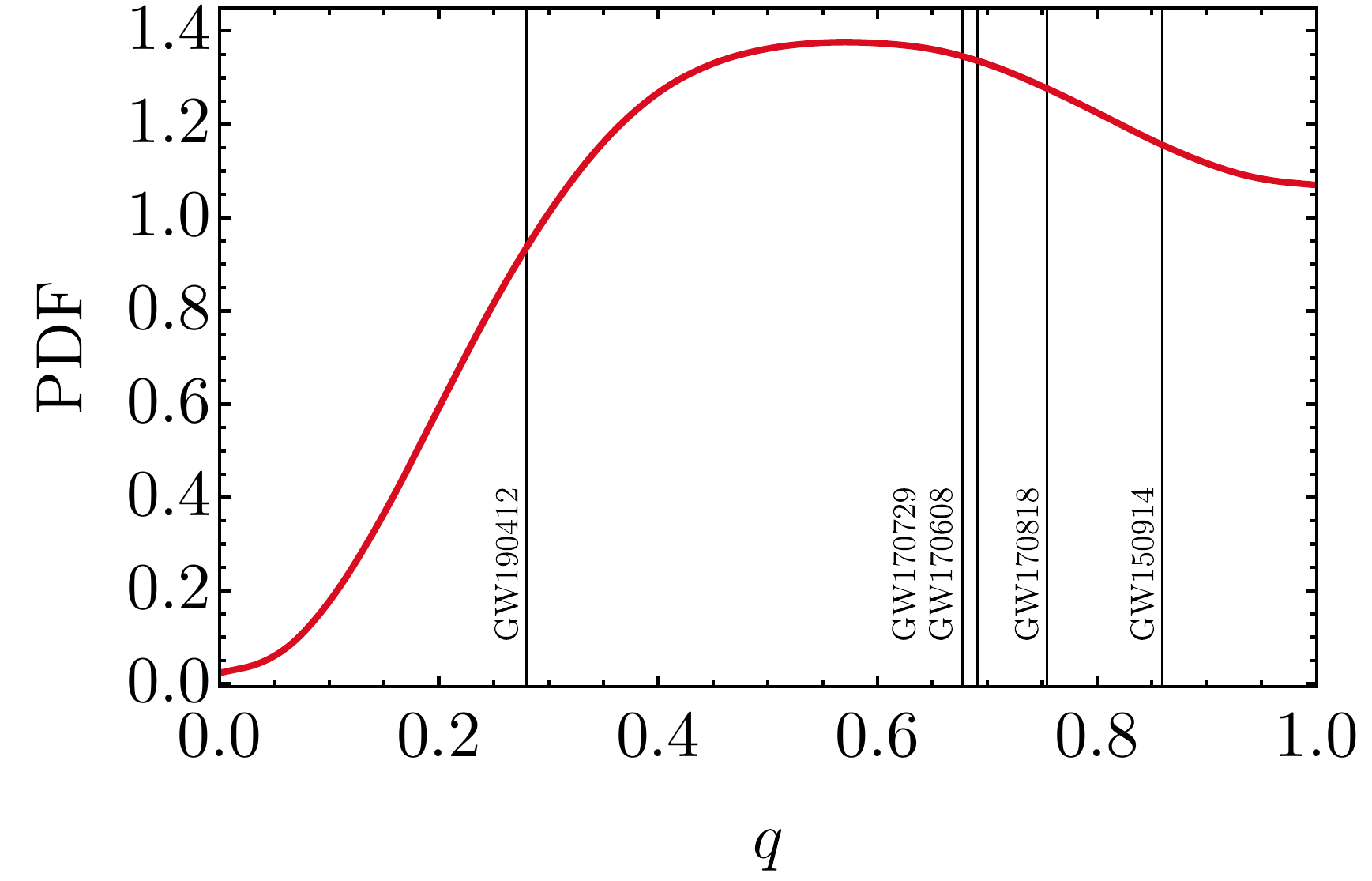}
  \includegraphics[width=0.49\textwidth]{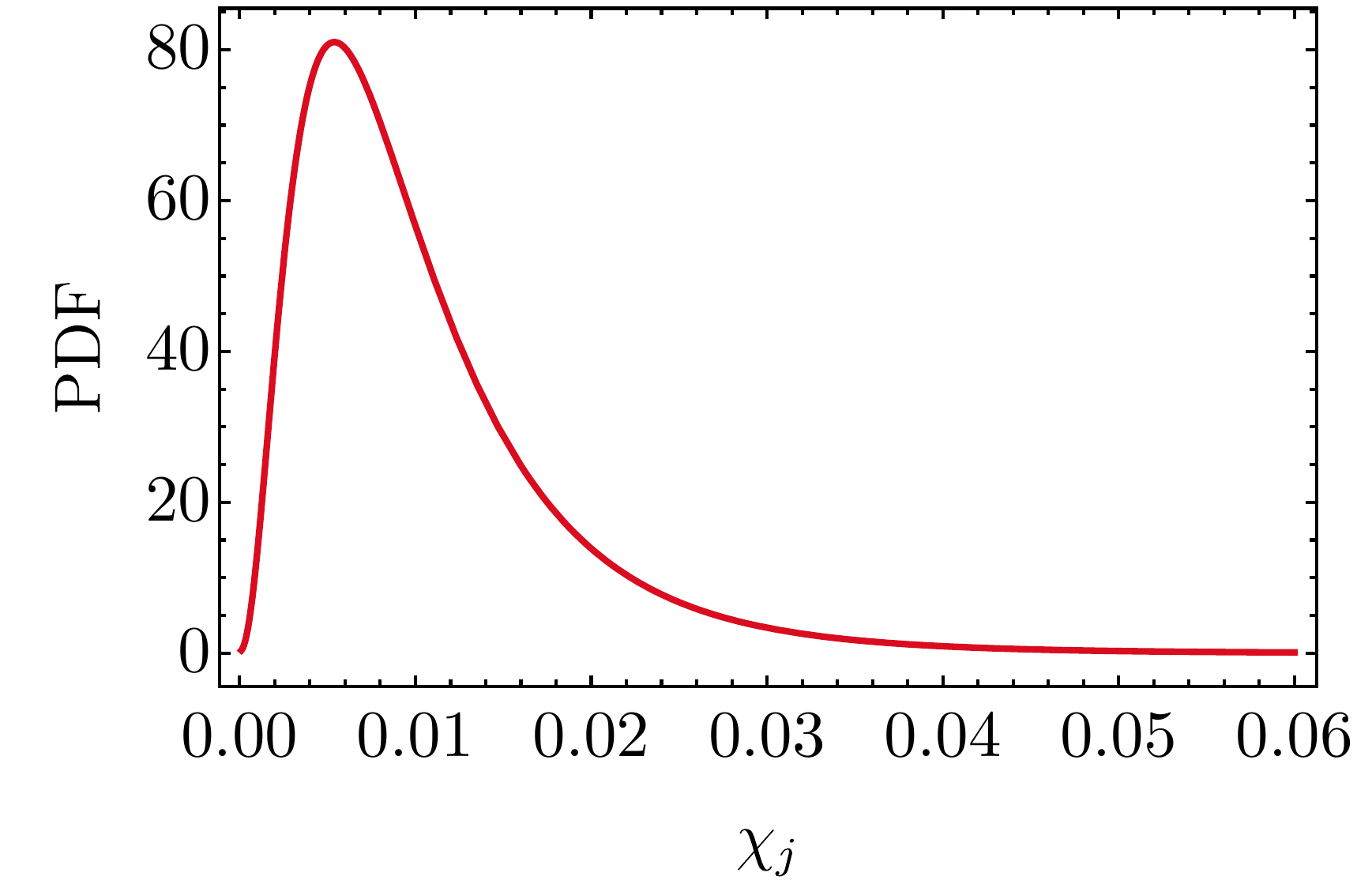}
  \includegraphics[width=0.49\textwidth]{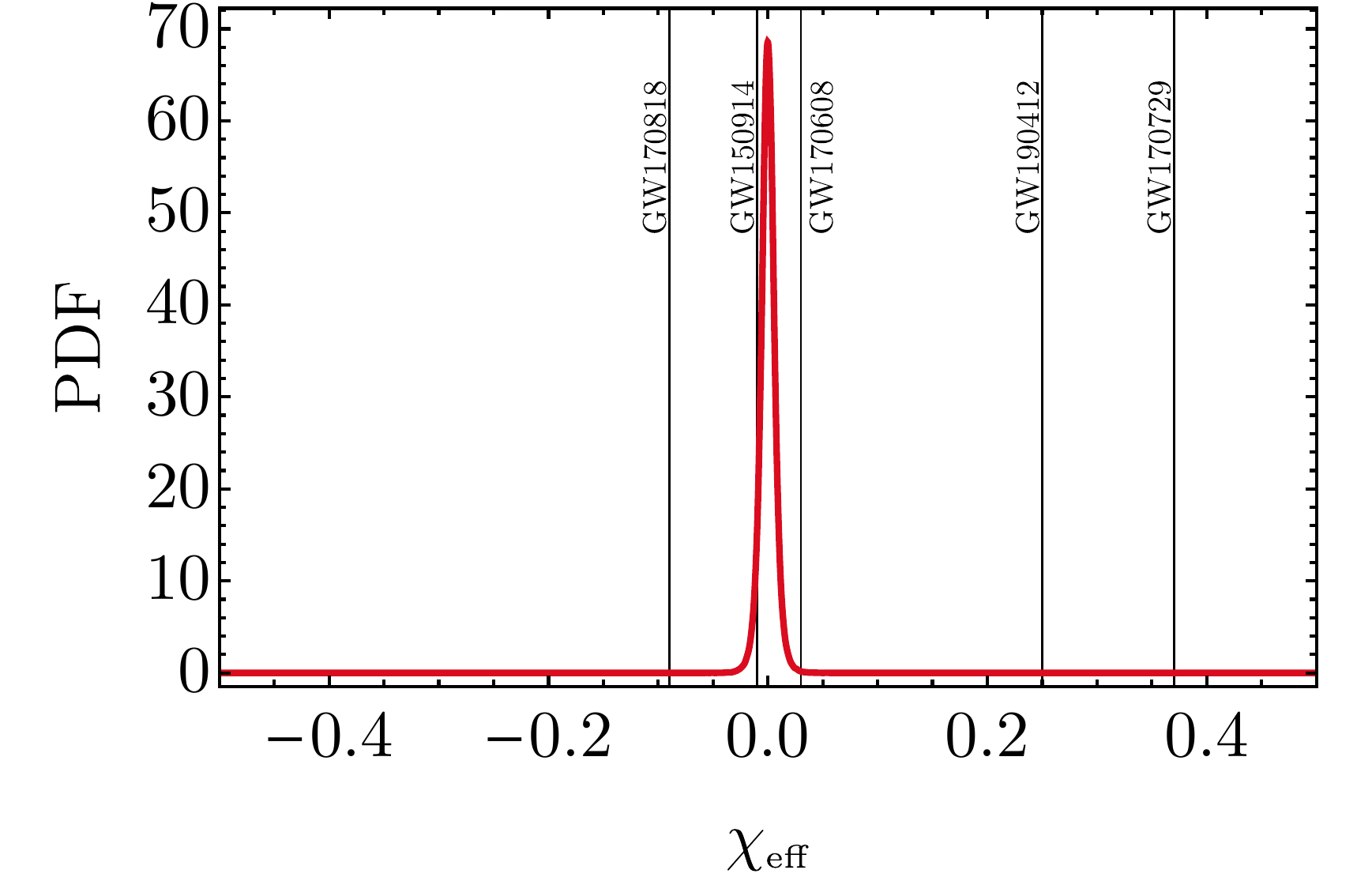}
 \caption{Prior distributions for the quantities $M_\text{\tiny chirp}$, $q$, $\chi_j$ ($j=1,2$) and $\chi_\text{\tiny eff}$, respectively, for the case of PBH binaries with no accretion. The best-fit values for the PBH mass function (Eq.~\eqref{psi}) are $M_c = 18 M_\odot$ and $\sigma =0.55$ \cite{DeLuca:2020qqa}. For illustrative purposes, we show also the central values for the  events analysed in the following (but obtained with the standard agnostic priors). Here we are showing the chirp mass values and distribution in the source-frame. 
}
 \label{fig:priors_no_acc}
 \end{figure*}

In Fig.~\ref{fig:priors_no_acc}, we show the distributions for the binary BH parameters at the formation, which coincide with the priors relevant for the parameter estimation in the case without accretion (or when accretion is inefficient). With the procedure previously outlined, the $\chi^2$-analysis using data from the O1-O2 runs as well as GW190412 selects the values $M_c = 18 M_\odot$ and $\sigma = 0.55$ for the initial mass function parameters~\cite{DeLuca:2020qqa}. The chirp mass distribution is peaked around the central scale of the PBH mass function, with a rapid fall-off at higher masses, while the distribution of the mass ratio is quite broad. Moreover, due to the fact that the PBH spins are negligible at formation, the distribution of the effective spin parameter is very narrow around its central value $\chi_\text{\tiny eff} \sim 0$. For a reference, in this and following plots we have superimposed the corresponding values for the GW events analyzed in this paper, see Table~\ref{tab:Log Bayes Factor} for details.

\begin{figure*}
  \includegraphics[width=0.49\textwidth]{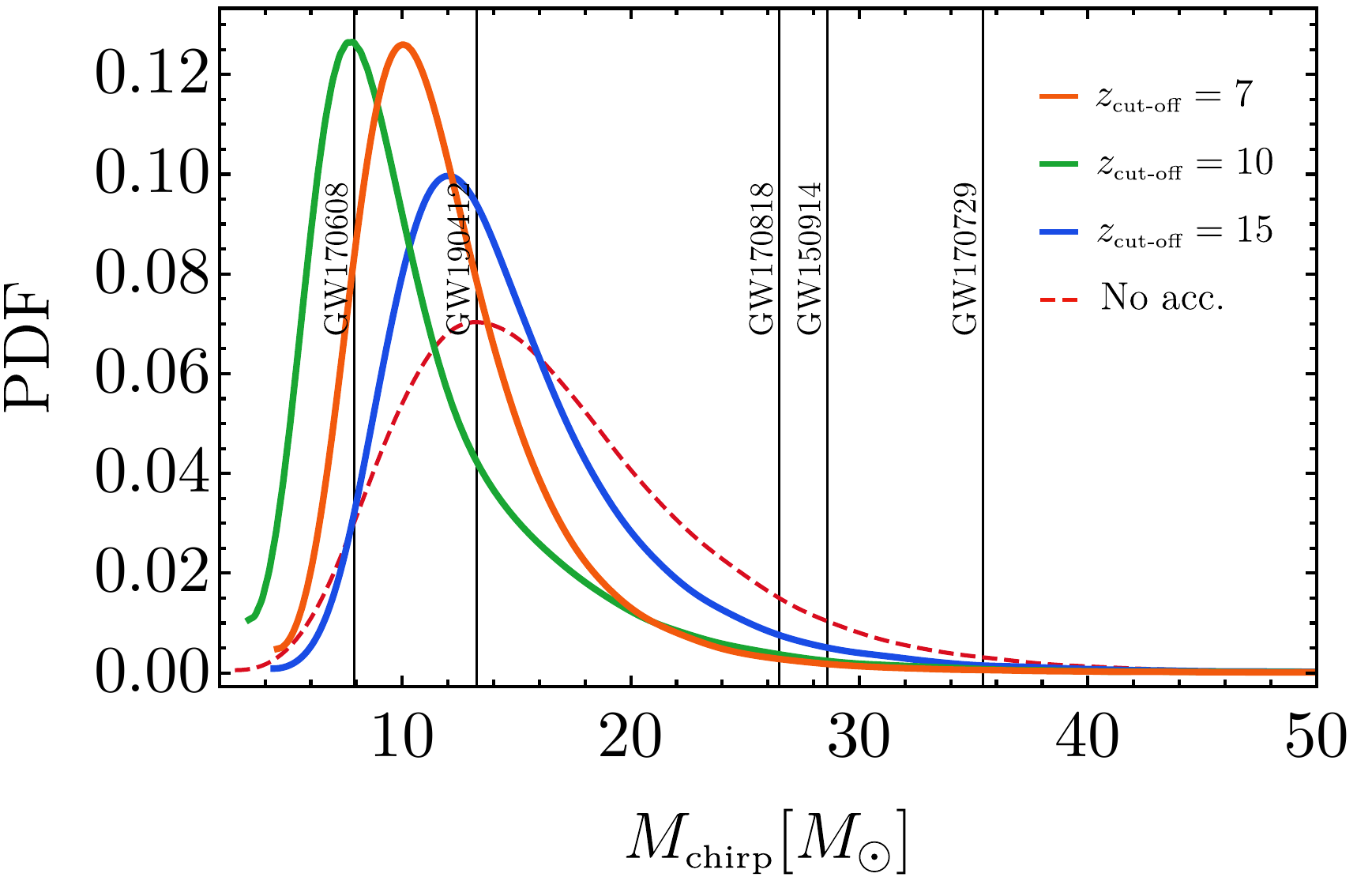}
  \includegraphics[width=0.49\textwidth]{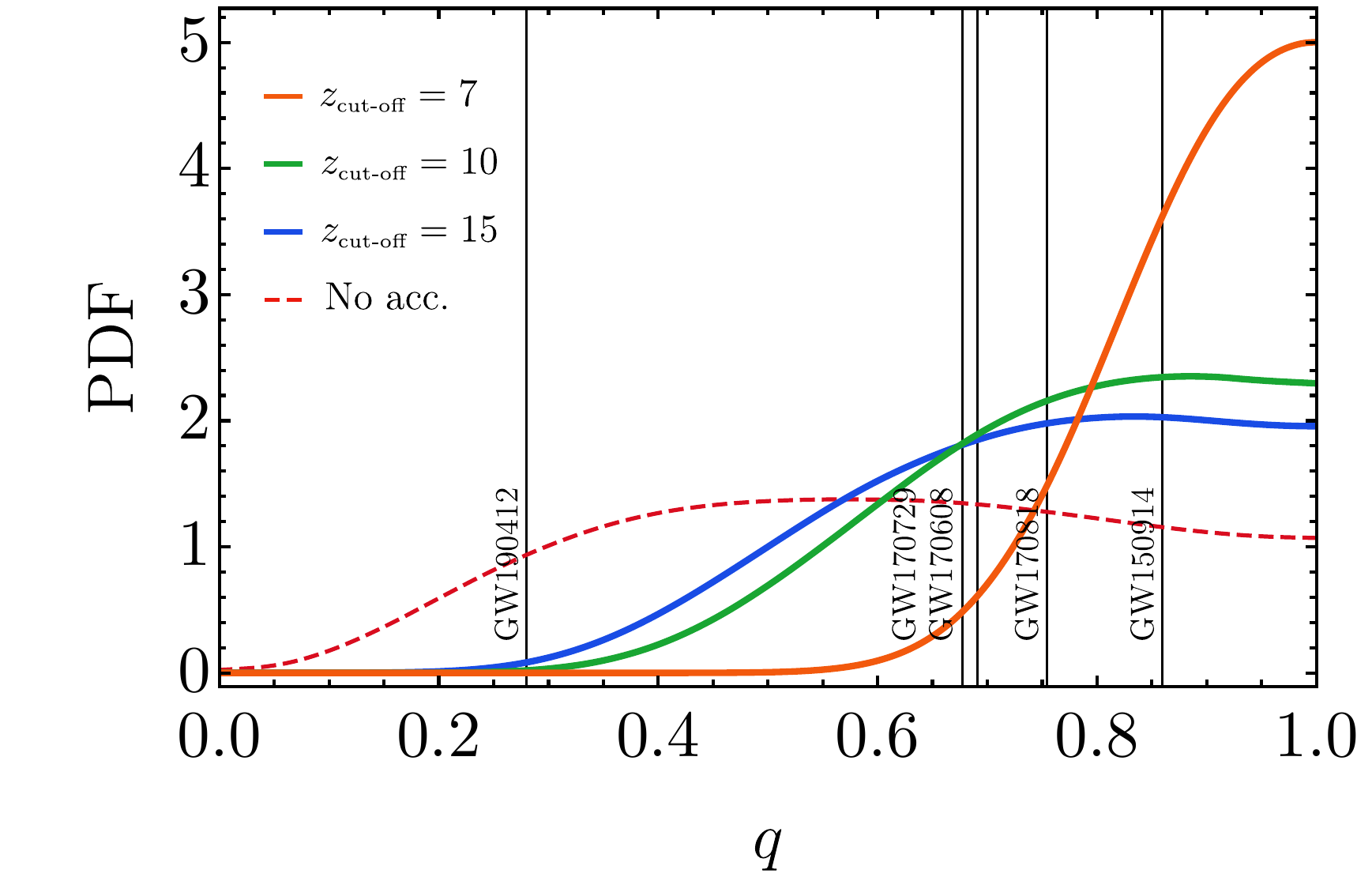}
  \includegraphics[width=0.49\textwidth]{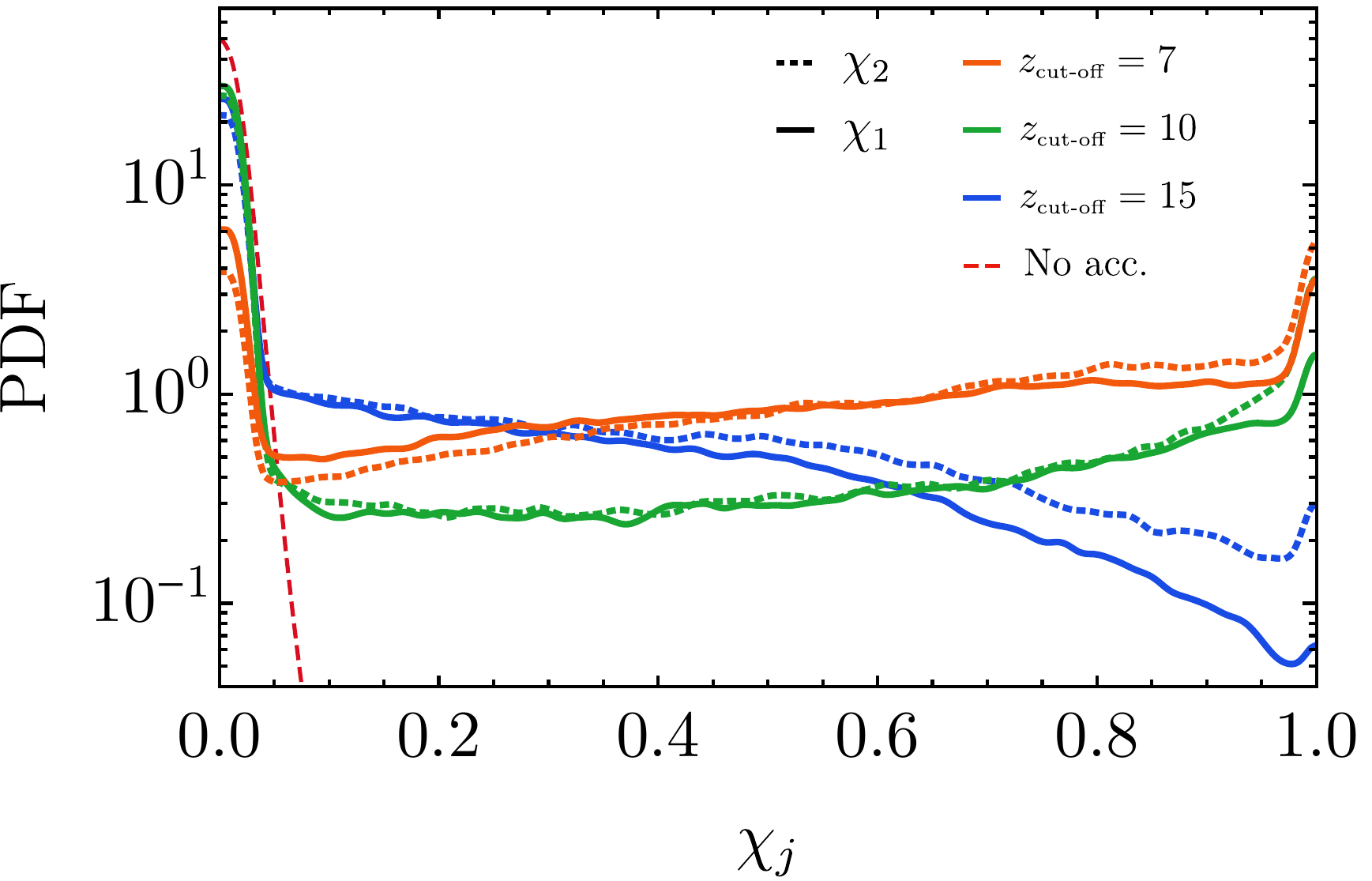}
  \includegraphics[width=0.49\textwidth]{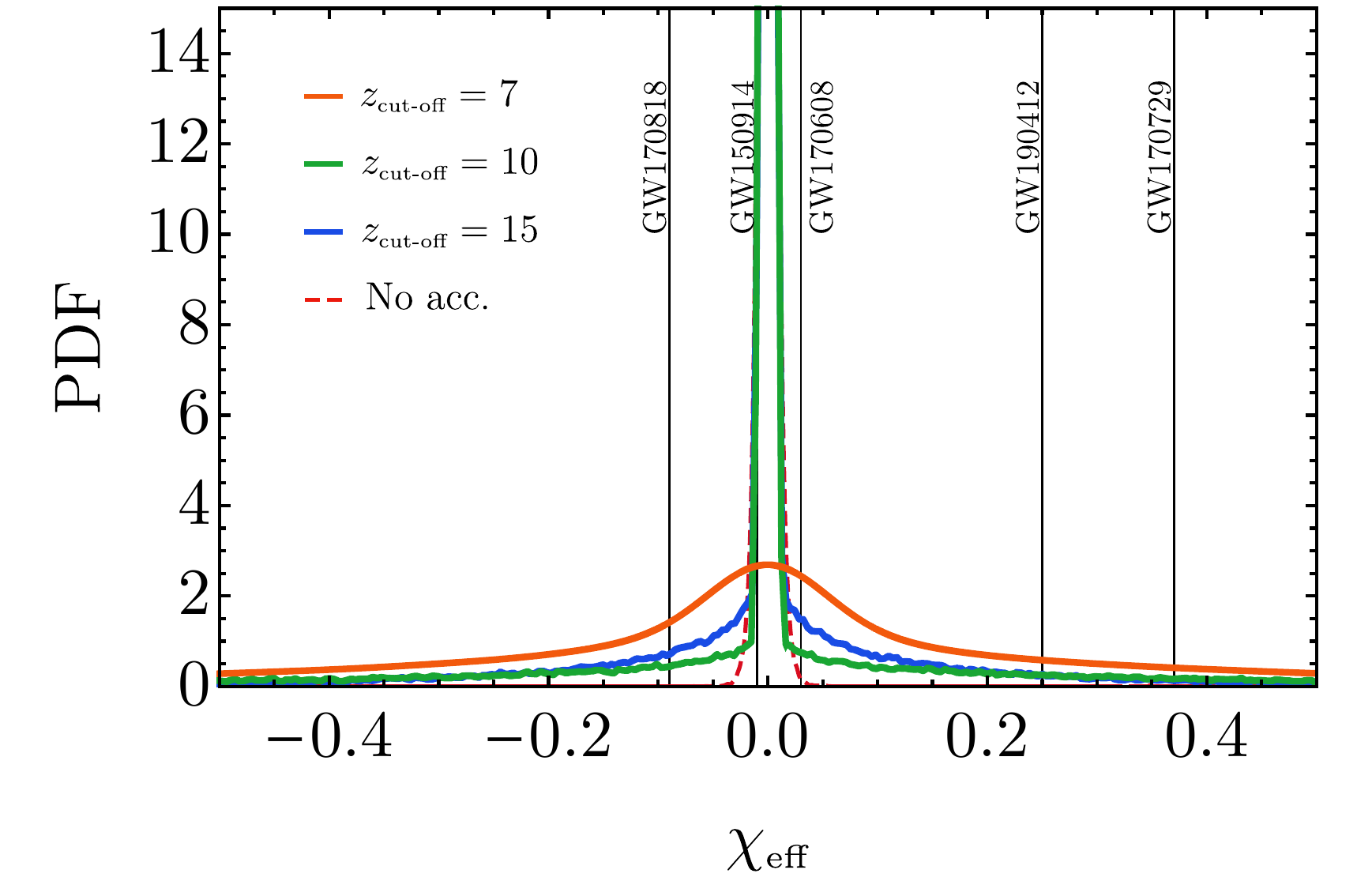}
 \caption{Same as Fig.~\ref{fig:priors_no_acc} for the case with accretion and $z_\text{\tiny cut-off}=\{ 15, 10, 7\}$, where 
 the PBH mass function is characterised by the set of parameters 
 $M_c = 14.1 M_\odot$ and $\sigma =0.32$, 
 $M_c = 8.7 M_\odot$ and $\sigma =0.29$, 
 $M_c = 7.3 M_\odot$ and $\sigma =0.15$,
 respectively
 \cite{DeLuca:2020qqa}. 
 As the priors have correlations between variables, the distribution shown for each individual observable is built by marginalising over all the remaining quantities.
 }
 \label{fig:priors_acc}
 \end{figure*}

In Fig.~\ref{fig:priors_acc} we show the priors distributions for the binary BH parameters for the case with accretion, assuming cut-off redshifts $z_\text{\tiny cut-off} = \{ 15, 10, 7 \}$. 
One can see that the chirp mass distribution gets narrower with respect to the case without accretion. The mass scale at which the distribution peaks is determined by two competitive effects: i)~the selection of the best-fit $M_c$ for the initial PBH mass function which, as accretion lasts longer, peaks at smaller values; ii)~the effect of the mass evolution which pushes the distribution to higher masses. 
Furthermore, as discussed above, the mass ratio distribution is skewed towards the fixed point $q=1$ due to the accretion effects onto the binary BHs.
Looking at the individual spin distributions one can appreciate that, as accretion takes place, they are also pushed towards unity, with the lighter binary component always spinning faster than the heavier one due to the characteristic behaviour of the individual mass accretion rates~\cite{DeLuca:2020qqa}. Furthermore, since accretion is inefficient for small masses and very efficient above a certain mass threshold, the marginalized spin distribution is bimodal with a peak at $\chi_{1,2}\approx0$ and another (smaller) peak towards extremality, which is anyway less relevant unless accretion is very strong.
Finally, the spread of the marginalized distribution of $\chi_\text{\tiny eff}$ around its central (zero) value
does not change monotonically as a function of the accretion. Generically, for very strong accretion ($z_\text{\tiny cut-off}\lesssim7$) the distribution tends to become broader. However, intermediate cases ($z_\text{\tiny cut-off}\approx 10$) are more involved, for example the marginalized distribution of $\chi_\text{\tiny eff}$ for $z_\text{\tiny cut-off}= 10$ is narrower than that for $z_\text{\tiny cut-off}= 15$ for which accretion is less relevant. The reason can be understood by looking at the individual spin distributions. Due to the aforementioned bimodal shape, it is more likely to obtain intermediate values of $\chi_{1,2}$ (and hence a broader distribution of $\chi_\text{\tiny eff}$) for $z_\text{\tiny cut-off}=15$ than for $z_\text{\tiny cut-off}=10$. 

\begin{figure*}
  \includegraphics[width=0.32\textwidth]{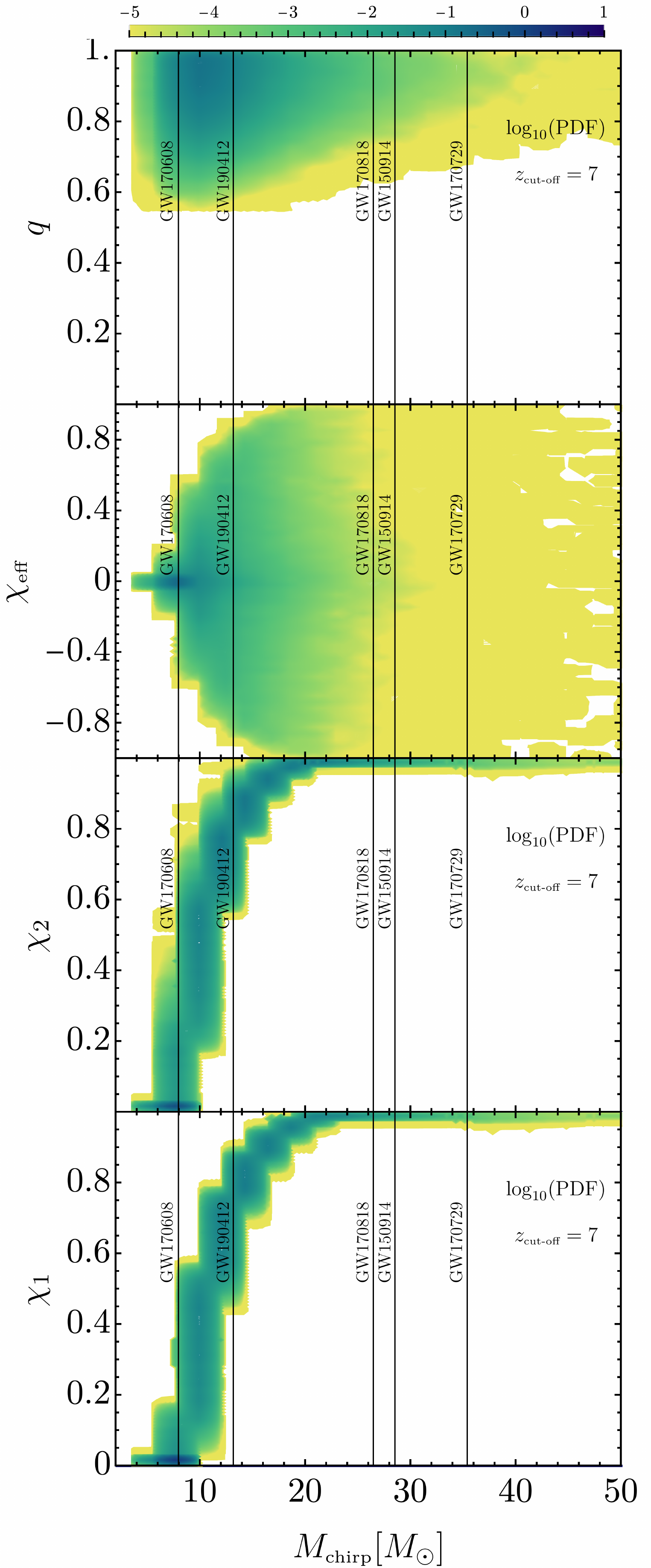}
  \includegraphics[width=0.32\textwidth]{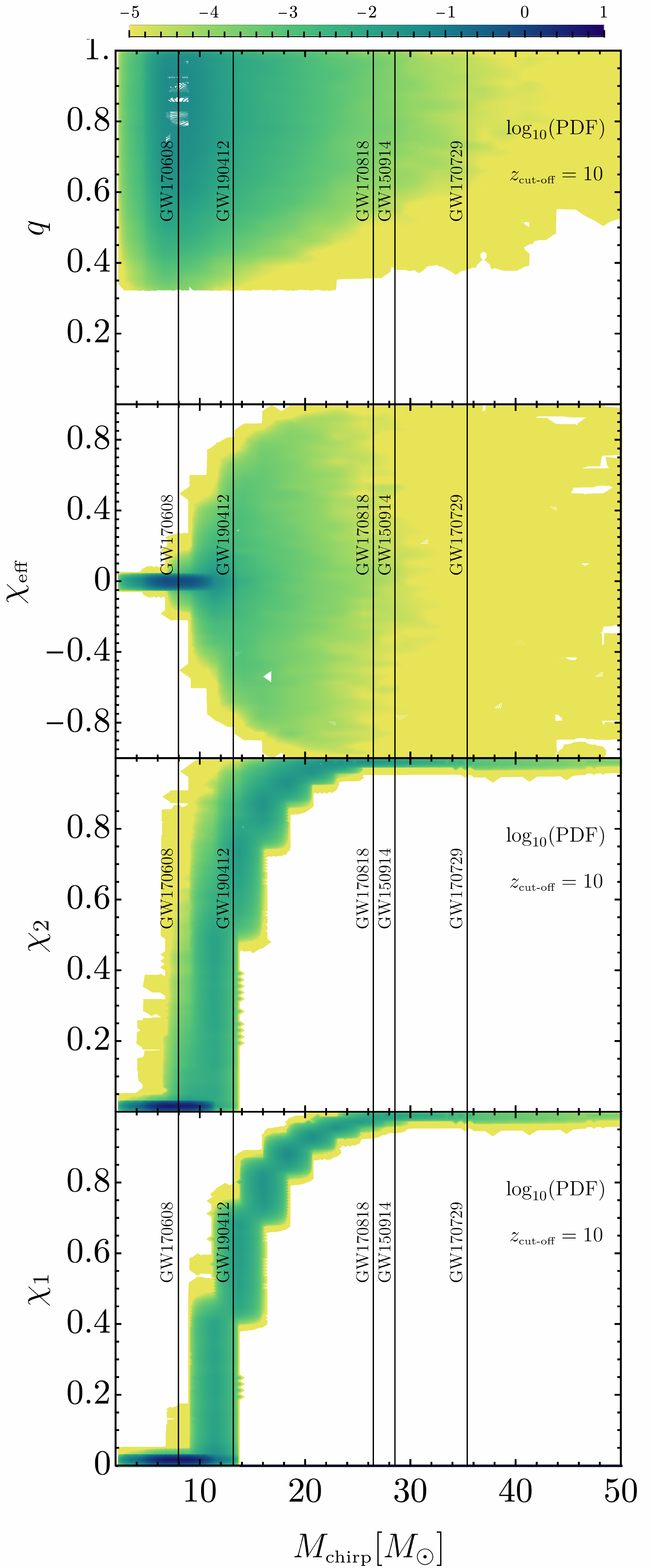}
  \includegraphics[width=0.32\textwidth]{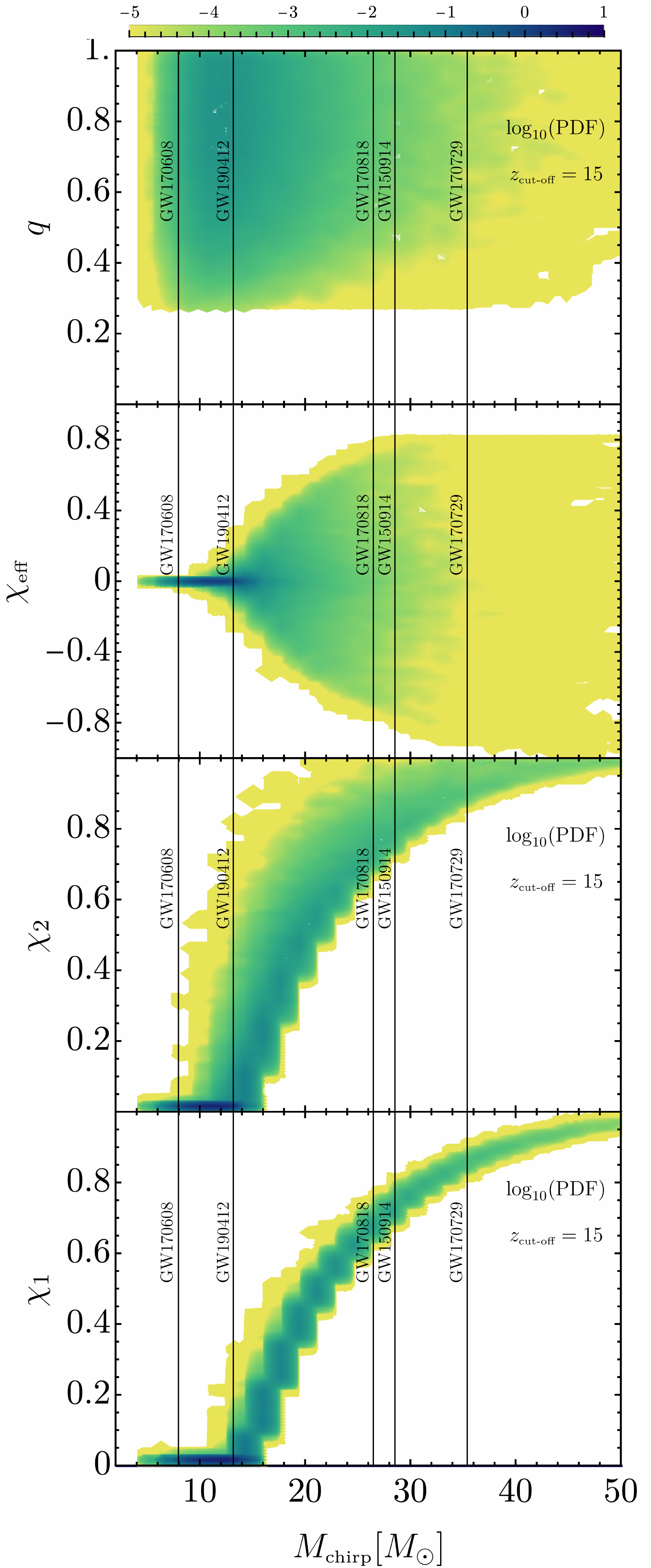}
 \caption{
 Prior distributions as in Fig.~\ref{fig:priors_acc} in the plane $(M_\text{\tiny chirp}, q)$, $(M_\text{\tiny chirp}, \chi_\text{\tiny eff})$, $(M_\text{\tiny chirp}, \chi_j)$ respectively, to show the correlation between quantities. In particular,  a strong positive correlation between chirp mass and individual spins is observed, which is also inherited by $\chi_\text{\tiny eff}$. White color identifies regions with negligible values of the PDF. 
 }
 \label{fig:priors_acc_2}
 \end{figure*}

We stress that the spin distributions shown in Fig.~\ref{fig:priors_acc} are marginalized over the masses. Indeed, an additional crucial difference between the priors without and with accretion lies in the correlation between different parameters. In an accreting system, the masses and the spins of the components are intertwined, leading to a strong correlation between large
chirp masses and high values of the spins. A similar, less pronounced, correlation is present between large spins (and hence a wide spread of the $\chi_\text{\tiny eff}$) and $q \sim 1$.
In brief, in the accreting PBH scenario high-mass binaries tend to be symmetrical and to have large spins (and hence a broad distribution of $\chi_\text{\tiny eff}$~\cite{DeLuca:2020fpg,DeLuca:2020qqa}). This is shown in Fig.~\ref{fig:priors_acc_2}, a comprehensive plot to which we shall often refer to when interpreting the results of the parameter estimation in Sec.~\ref{sec:results}.

\section{Parameter estimation for coalescing BH binaries}\label{sec:analysis}
Bayesian parameter estimation is the conventional technique used to determine the posterior distributions  $p(\tth|\ddat,\mdl)$  of parameters $\tth$ in a model $\mdl$ from the observed data $\ddat$. A key feature of this framework is that it folds in the expectations of distribution of the parameter values via a probability density function called the \textit{prior},  $\Pi =\pp(\tth|\mdl)$. The observation itself contributes to the inferred posterior distribution via a \textit{likelihood function} $\pp(\ddat|\tth,\mdl)$. The Bayes theorem  states that the posterior probability density function $\pp(\tth|\ddat,\mdl)$ for the parameters $\tth$ is given by~\cite{BayesBooks1}
\begin{align}
\label{eq:bayes_thrm}
\pp(\tth|\ddat,\mdl) = \frac{\pp(\tth|\mdl) \pp(\ddat|\tth,\mdl)}{Z},
\end{align}
where $\ddat= \mdl+\nn$, $\nn$ is the noise, and the normalizing factor $Z= \pp(\ddat|\mdl)$ is called the {\it evidence}. 
Assuming the noise model is Gaussian and stationary, the likelihood function can be expressed as  
\begin{align}
\label{eq:likelihood}
L = \pp(\ddat|\tth,\mdl) \propto e^{-\frac{1}{2} \left\langle \nn|\nn  \right\rangle}= e^{-\frac{1}{2} \left\langle \ddat - \mdl|\ddat - \mdl  \right\rangle},
\end{align}
where $\left\langle .|. \right\rangle$ denotes the weighted inner product. For our case, $\mdl$ is the frequency-domain binary BH gravitational waveform $\widetilde{h(f)}$,  $\ddat$ is the output of the GW interferometers expressed in the frequency domain. Further, 
\begin{equation}
\left\langle \mdl|\ddat \right\rangle = \int_{0}^{\infty} df\, \frac{\widetilde{h(f)}^{*} \times \widetilde{\ddat (f)}  }{S_{n}(f)},
\end{equation}
where $S_{n}(f)$ is the power spectral density (PSD) of the instrument.      
While the likelihood reflects the tendency of the data, the \textit{prior} $\pp(\tth|\mdl)$ encapsulates our expectations and is a choice that is made in the inference. 
Furthermore, the evidence $Z$ associated with the different choices of priors quantifies which of the prior choices is favoured by the observed data. The evidence is obtained by completely marginalizing the posterior and can be written as  
\begin{align}
Z= \pp(\ddat|\mdl) = \int d \tth L(\ddat|\tth) \times \Pi(\tth)\,.
\end{align}
The evidence takes into account the goodness of fit of the data with the model of the signal through the $ L(\ddat|\tth)$ and the volume of the prior space $\Pi(\tth)$, thereby penalizing over-fitting. From a given data, if the inference is made using two or more models of the signal or the prior distributions, then the ratios of the evidences known as the Bayes factor is used to compare the performance of these assumptions.
Note that this interpretation inherently assumes that each of the prior distribution is equally likely before conducting the experiment. 

\subsection{Details of the parameter estimation setup}
\label{subsec: Bilby setup}

We use the open science data for the binary BH coalescences as outlined in Ref.~\cite{Romero-Shaw:2020owr}, using data from 
the two LIGO interferometers. We use the {\it BILBY} Bayesian inference library~\cite{Ashton:2018jfp,Romero-Shaw:2020owr} to perform full Bayesian parameter estimations and infer all the 15 parameters of the GW coalescence waveform model, namely the source-frame masses ($M_1$, $M_2$), the dimensionless spins magnitudes ($\chi_1$, $\chi_2$) of the BHs, 4 angle variables ($\theta_{1,2}$, $\delta \phi$, $\delta_{JL}$) that describe the BH spin directions, the inclination angle ($\iota$), the polarization angle ($\psi$), the phase at coalescence ($\phi_c$), the time of coalescence ($t_c$), the right ascension ($\alpha$), the declination ($\delta$) and the luminosity distance ($d_L$). 
We use a dynamic nested sampler implementation called DYNESTY \cite{Speagle, Skilling2,Skilling1} within the BILBY package. The sampler is run with 1000 live points and 100 walks to produce the posterior distribution presented in Sec. \ref{sec:results}. 

For the analysis of all events under consideration, excluding GW190412, we use only the dominant $l=m=2$ harmonic. On the other hand, for GW190412 (the most asymmetric binary BH system included in this analysis) the subdominant $l=m=3$ harmonic has been measured~\cite{LIGOScientific:2020stg}). Therefore, for this event we also included higher harmonics in the waveform model and we shall discuss the results of the two cases (with and without higher harmonics) separately. Indeed, while the inclusion of higher harmonics does not change the inferred parameter values for GW190412 significantly~\cite{collaboration2020gw190412}, it might change the evidence of a given model. Furthermore, the impact of higher harmonics on the posterior distributions of GW190412 has been demonstrated only using agnostic priors, while the case of PBH-motivated priors requires an independent investigation.
When considering only the dominant $l=m=2$ harmonic, we used the 'IMRPhenomPv2' frequency domain waveform implementation from LALSuite~\cite{lalsuite} for the likelihood computations. 
For the analysis using higher harmonics, instead, we used the `IMRPhenomHM' waveform approximant~\cite{Kalaghatgi:2019log}, including $(l,m) =\{(2, 2), (2, 1), (3, 3), (3, 2), (4, 4), (4, 3)\}$ in the parameter estimation. Notice that this model assumes spin vectors orthogonal to the orbital plane, so in this case the spin angles are removed from the parameter estimation, and the number of parameters is smaller than in the case of the dominant-mode only, due the difference in the waveform model.
The priors are given on $ \{M_1, M_2,\chi_{1}^{||},\chi_{2}^{||},d_L,\alpha,\delta,\psi,\phi_c, \iota, t_c\}$, where $\chi_{i}^{||}\equiv \chi_i$ is the only non-vanishing components of the spin of the $i$-th BH, which are parallel to the orbital angular momentum.

The PSD of the detectors is calculated from the data using the GWpy package \cite{Walker_2018,PhysRevD.99.082002}.  A median value of PSD is used for the computation of the likelihood. 

Parameter estimation is carried out for all the 15 (reduced to 11 when higher harmonics are included) binary BH parameters but we present only the relevant parameters in view of clarity. As explained in the previous section, the assumption of BHs having a primordial origin modifies the priors on the masses and spin magnitudes of the BHs. The implementation of these priors in the parameter estimation setup is described in Secs.~\ref{sec:proper}, and \ref{sec:accr}. For all other parameters, we choose the standard non-informative priors as summarized in Table~\ref{tab:extrinsic-prior}.

\begin{table}[]
\begin{tabular}{|c|c|}
\hline
\textbf{Parameter}                      & \textbf{Prior}                                      \\ \hline
$\theta_{i, j}$                 & Sin prior in $[0, \pi]$                     \\ \hline
$\delta \phi$                   & Uniform prior in $[0, 2 \pi]$               \\ \hline
\ $\delta_{JL}$ & Uniform  prior in $[0, 2 \pi]$               \\ \hline
$t_c$                           & Uniform prior around the event trigger time \\ \hline
$\phi_c$                        & Uniform in $[0, 2 \pi]$                     \\ \hline
$\alpha$                            & Uniform prior in $[0, 2 \pi]$               \\ \hline
$\delta$                           & Cosine prior in $[- \pi/2, \pi/2]$          \\ \hline
$d_L$                           & Power law prior in $[50,2000]$ Mpc          \\ \hline
$\iota$                         & Sin prior in $[0, \pi]$                     \\ \hline
$\psi$                          & Uniform in $[0, 2 \pi]$                     \\ \hline
\end{tabular}
\caption{List of the standard, non-informative priors we use for 11 (out of 15) waveform parameters. The PBH-motivated priors on the BH masses and spins are summarized in Table~\ref{tab:Prior-tab}.}
\label{tab:extrinsic-prior}
\end{table}

\subsection{Parameter estimation with PBH-motivated priors: non-accreting case}
\label{sec:proper}

If accretion during the cosmological evolution of PBHs is not efficient, their masses and spins are those at formation. As previously discussed, for the former we use the log-normal distribution given in Eq.~\eqref{psi} with parameters $M_c$ and $\sigma$ obtained by a population-driven analysis as discussed in Sec.~\ref{sec2.3}, assuming all O1+O2+GW190412 events detected so far are of primordial origin~\cite{DeLuca:2020qqa}. The real distribution for the spins is complicated~\cite{DeLuca:2019buf} and we approximate it by a Gaussian with central value $\mu=0.005$ and standard deviation $\sigma=0.002$. In our analysis, we have checked that the specific form of the distribution is irrelevant as long as it only has effective support for $\chi_{1,2}\lesssim0.01$, as predicted by the formation scenario~\cite{DeLuca:2019buf}. A comparison between the PBH-motivated priors and the agnostic ones adopted by the LVC analysis is presented in Table~\ref{tab:Prior-tab}.
Finally, since the spin orientations are unknown in most scenarios, in all cases we assume that the spin vectors are isotropically distributed.

\begin{table*}[]
\begin{tabular}{c|c|c|}
\cline{2-3}
& {\bf PBH-motivated priors (no accretion)}   & {\bf Standard (LVC) Priors}                            \\ \hline
\multicolumn{1}{|c|}{$M_1$} & Lognormal with $M_c = 18 M_\odot$ and $\sigma=0.55$ & Uniform, $M_1\in[3,30]M_\odot$ \\ \hline
\multicolumn{1}{|c|}{$M_2$}  & Lognormal with $M_c = 18 M_\odot$ and $\sigma=0.55$ & Uniform, $M_2\in[3,30]M_\odot$ \\ \hline
\multicolumn{1}{|c|}{$\chi_1$} & Gaussian with $\mu = 0.005$ and $\sigma=0.002$ & Uniform, $\chi_1\in[0,0.99]$                        \\ \hline
\multicolumn{1}{|c|}{$\chi_2$} & Gaussian with $\mu=0.005$ and $\sigma=0.002$   & Uniform, $\chi_2\in[0,0.99]$                        \\ \hline
\end{tabular}
\caption{Summary of the standard priors adopted by the LVC analysis and those motivated by a PBH scenario without accretion for the binary masses and spin. The lognormal distribution for the PBH masses is defined in Eq.~\eqref{psi}, whereas we approximate the spin distribution by a Gaussian with central value $\mu$ and spread $\sigma$.}
\label{tab:Prior-tab}
\end{table*}

\subsection{Parameter estimation with PBH-motivated priors: accreting case}
\label{sec:accr}

As discussed in Sec.~\ref{sec:priors_accr}, if accretion is relevant during the cosmological evolution of the PBHs, the masses and the spins of the BHs evolve. As a consequence, the distributions of the masses and the spins of the BHs observed by the detectors are correlated when the binary BH enters the LIGO-Virgo band. 

When accretion is significant, one would ideally need to implement  {\it conditional} priors. In other words, unlike the cases of non-accreting PBH priors or agnostic priors, the prior distribution $\Psi=\Psi(M_1,M_2,\chi_1,\chi_2)$ for accreting PBHs cannot be factorized as $\Psi=\Psi_a(M_1)\Psi_b(M_2)\Psi_c(\chi_1)\Psi_d(\chi_2)$. 

However, the practical implementation of conditional priors is challenging and involves handling multi-dimensional interpolated prior functions.  Most nested samplers implementation involve taking the inverse cumulative distribution function of the prior distribution and this can be complicated for a generic parameter correlation. This is especially a problem if the multidimensional priors have sharp features, which is our case in some parts of the parameter space.  Finally, the distribution $\Psi(M_1,M_2,\chi_1,\chi_2)$ is not known in an analytical form and have to be tabulated numerically.

To overcome these difficulties, in the case of accreting PBHs we have adopted the following approximate procedure. 
For a given accretion cut-off $z_\co$, we build a Monte Carlo simulation of the PBH population, from which we measure the exact mass function in tabulated form. The latter is independent of the values of the spins and of other parameters. Then, we approximate the numerical mass function with the  log-normal distribution given in Eq.~\eqref{psi}, with parameters $M_c$ and $\sigma$ obtained by fitting (see Table~\ref{tab:mass-acc-prior}).

\begin{figure*}
\includegraphics[width=0.5\textwidth]{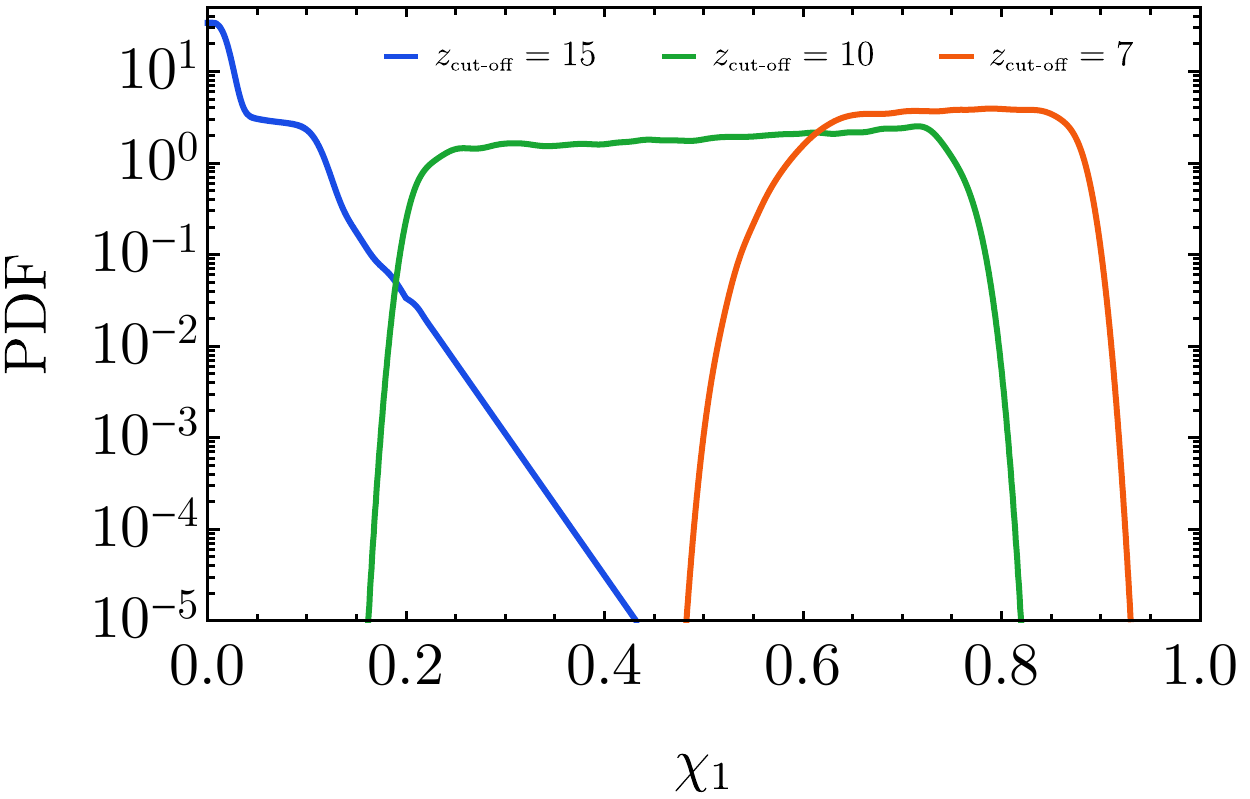}
\includegraphics[width=0.5\textwidth]{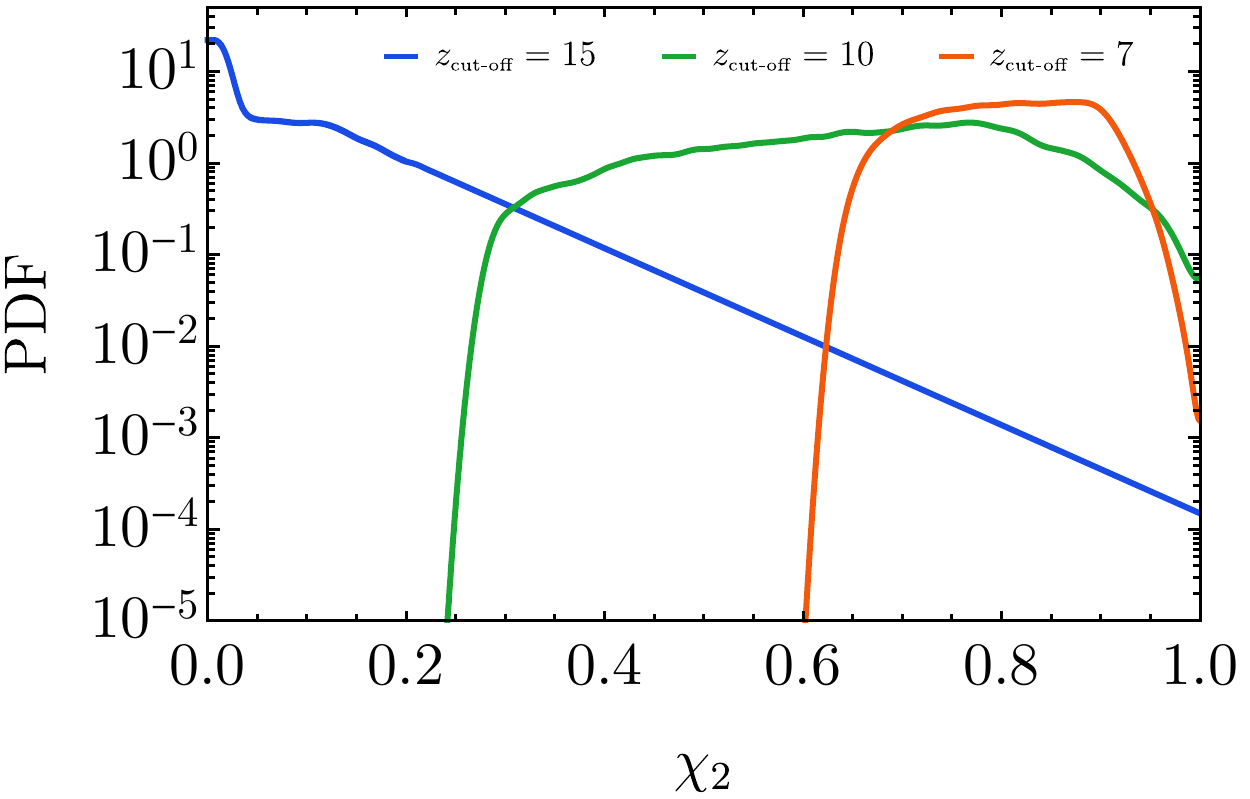}
\caption{Prior distributions for PBH spins, for GW190412 parameter estimation, with significant accretion for $z_\co$ = 7, 10, 15.}
\label{fig:spin-acc-prior}
\end{figure*}

To construct manageable spin priors in a factorized form, for each event, we accounted for the  correlation  of $\chi_{1,2}$ with the masses by building their approximate distribution considering only the events in the simulation constrained to have $M_\text{\tiny chirp}$ within the corresponding GW event chirp mass $90\%$ C.I. (as inferred with standard priors).
We expect this to be accurate since the chirp mass, when performing the parameter estimation, does not correlate strongly with other parameters and is indeed much less sensitive to the choice of different priors.
Thus, our approximation should be valid as long as the value of $M_\text{\tiny chirp}$ inferred by using PBH-motivated priors does not differ significantly from that inferred by using agnostic priors, as can be checked a posteriori. The spin priors obtained through this procedure are shown in Fig.~\ref{fig:spin-acc-prior} for the representative case of GW190412. We use these numerical tabulated priors in our analysis of the accreting PBH scenario.

\begin{table}[]
\begin{tabular}{|c|c|c|c|}
\hline
\textbf{Parameter} &
  $z_\text{\tiny cut-off} = 7$ &
  $z_\text{\tiny cut-off} =10$ &
  $z_\text{\tiny cut-off} =15$ \\ \hline
$M_{1}$ &
  \begin{tabular}[c]{@{}c@{}}$M_{c} = 15.78 \Msun$\\ $\sigma = 0.109$\end{tabular} &
  \begin{tabular}[c]{@{}c@{}}$M_{c} = 17.03 \Msun$\\ $\sigma = 0.138$\end{tabular} &
  \begin{tabular}[c]{@{}c@{}}$M_{c} = 17.56 \Msun$\\ $\sigma = 0.152$\end{tabular} \\ \hline
$M_{2}$ &
  \begin{tabular}[c]{@{}c@{}}$M_{c} = 14.122 \Msun$\\ $\sigma = 0.109$\end{tabular} &
  \begin{tabular}[c]{@{}c@{}}$M_{c} = 13.31 \Msun$\\ $\sigma = 0.138$\end{tabular} &
  \begin{tabular}[c]{@{}c@{}}$M_{c} = 13.197 \Msun$\\ $\sigma = 0.152$\end{tabular} \\ \hline
\end{tabular}
\caption{Parameters of the lognormal prior distribution for PBH masses $M_{1}$ and $M_{2}$ with significant accretion for the representative cases $z_\co$ = 7, 10 and 15. }
\label{tab:mass-acc-prior}
\end{table}

\section{Results}
\label{sec:results}
In this section we present the results of the parameter estimation for some representative GW events by implementing the PBH-motivated priors discussed in Sec.~\ref{sec:priors} in the BILBY infrastructure~\cite{Ashton:2018jfp,Romero-Shaw:2020owr} and performing the analyses presented in Sec.~\ref{sec:analysis}. We shall always quote binary's masses in the source frame. We discuss the scenarios with and without accretion separately.

\subsection{Without accretion}\label{sec:resnoaccr}

We divide the events thus far detected by the LVC~\cite{LIGOScientific:2018mvr,LIGOScientific:2020stg} into 4 categories and study one event in each category for the case of priors corresponding to PBHs without accretion as well as for the  standard agnostic priors. In particular, we consider (see Table~\ref{tab:Log Bayes Factor}):
\begin{enumerate}
    \item GW150914~\cite{Abbott:2016blz}, as representative for moderate mass, symmetrical, binary BH systems; %
    \item GW170608~\cite{Abbott:2017gyy}, as representative for low mass, symmetrical, binary BH systems;
    \item GW170729~\cite{LIGOScientific:2018mvr}, as representative for moderate mass, asymmetrical, binary BH systems;
    \item GW190412*~\cite{LIGOScientific:2020stg}, as representative for low mass, asymmetrical, binary BH systems. As for the previous sources, here in the analysis we considered only the dominant $l=m=2$ harmonic;
    \item GW190412~\cite{LIGOScientific:2020stg}, same as the previous entry, but including also higher harmonics in the waveform model, as previously explained.
\end{enumerate}
We distinguish between low-mass and moderate-mass binaries (depending on whether $M_{1,2}\lesssim 20 M_\odot$), between symmetrical and asymmetrical systems (depending on whether $q$ is compatible with unity, at least marginally within the errors, or not), and between non-spinning and spinning binaries (depending on whether the posterior distribution of $\chi_\text{\tiny eff}$ is compatible with zero or not). We stress that these parameters refer to the original ones obtained with the standard priors on the masses and  the spins adopted by the LVC (see Table~\ref{tab:Prior-tab}).
Further, since GW150914 was a very loud detection, we also study GW170818 to better understand the effect of the priors for a less loud event belonging to the first category.

\begin{table}[]
\tiny
\begin{tabular}{|c|c|c|c|c|c|c|c|}
\cline{1-8}
                               \multicolumn{2}{|c|}{}       
                            & \textbf{GW150914} & \textbf{GW170818} & \textbf{GW170608}  & \textbf{GW170729}   & \textbf{GW190412*} & \textbf{GW190412}\\ \hline
   \multicolumn{2}{|c|}{}  &
  \begin{tabular}[c]{@{}c@{}}Moderate mass,\\  Symmetrical,\\Non-spinning\end{tabular} &
  \begin{tabular}[c]{@{}c@{}}Moderate mass, \\ Symmetrical,\\Non-spinning\end{tabular} &
  \begin{tabular}[c]{@{}c@{}}Low mass,\\ Symmetrical,\\Non-spinning\end{tabular} &
  \begin{tabular}[c]{@{}c@{}}Moderate mass,\\ Asymmetrical,\\Spinning~(?)\end{tabular}  &
  \begin{tabular}[c]{@{}c@{}}Low mass, \\ Asymmetrical,\\Spinning\end{tabular} &
  \begin{tabular}[c]{@{}c@{}}Low mass, \\ Asymmetrical,\\Spinning\end{tabular}\\ 
  \hline
  \textbf{Param.} & \textbf{Prior} & \multicolumn{6}{|c|}{} \\
  \hline
   $M_{\rm chirp}$ &\begin{tabular}[c]{@{}c@{}}Standard  \\ PBH  \end{tabular}    & \begin{tabular}[c]{@{}c@{}} $27.92^{+1.55}_{-1.37}$ \\ $28.51^{+1.01}_{-0.88}$ \end{tabular}
  & \begin{tabular}[c]{@{}c@{}} $25.79^{+2.75}_{-2.26}$ \\ $25.90^{+2.45}_{-1.78}$ \end{tabular}   & \begin{tabular}[c]{@{}c@{}} $7.95^{+0.16}_{-0.17}$ \\ $7.90^{+0.21}_{-0.19}$ \end{tabular}       & \begin{tabular}[c]{@{}c@{}} $39.44^{+7.43}_{-7.00}$ \\ $32.90^{+4.18}_{-3.29}$ \end{tabular}   & \begin{tabular}[c]{@{}c@{}} $13.05^{+0.70}_{-0.41}$ \\ $12.94^{+0.87}_{-0.48}$ \end{tabular} &
  \begin{tabular}[c]{@{}c@{}} $13.43^{+0.68}_{-0.49}$ \\ $13.22^{+0.65}_{-0.53}$ \end{tabular}
\\
  \hline
    $M_1$ &\begin{tabular}[c]{@{}c@{}}Standard  \\ PBH  \end{tabular}    & \begin{tabular}[c]{@{}c@{}} $34.66^{+4.77}_{-2.66}$ \\ $35.43^{+4.12}_{-2.55}$ \end{tabular}
  & \begin{tabular}[c]{@{}c@{}} $35.15^{+8.87}_{-5.37}$ \\ $34.31^{+7.75}_{-4.61}$ \end{tabular}   & \begin{tabular}[c]{@{}c@{}} $11.49^{+4.02}_{-2.02}$ \\ $9.53^{+0.80}_{-0.48}$ \end{tabular}       & \begin{tabular}[c]{@{}c@{}} $58.61^{+14.88}_{-11.97}$ \\ $56.06^{+11.15}_{-12.07}$ \end{tabular}   & \begin{tabular}[c]{@{}c@{}} $28.90^{+5.15}_{-5.24}$ \\ $18.87^{+3.26}_{-3.37}$ \end{tabular}  &
  \begin{tabular}[c]{@{}c@{}} $28.55^{+3.17}_{-3.04}$ \\ $22.75^{+1.88}_{-2.43}$ \end{tabular} \\

$M_2$ &\begin{tabular}[c]{@{}c@{}}Standard  \\ PBH  \end{tabular}    & \begin{tabular}[c]{@{}c@{}} $29.86^{+2.87}_{-4.19}$ \\ $30.40^{+2.87}_{-4.19}$ \end{tabular}
  & \begin{tabular}[c]{@{}c@{}} $25.30^{+5.17}_{-5.99}$ \\ $26.18^{+4.42}_{-5.54}$ \end{tabular}   & \begin{tabular}[c]{@{}c@{}} $7.35^{+1.43}_{-1.71}$ \\ $8.65^{+0.45}_{-0.68}$ \end{tabular}       & \begin{tabular}[c]{@{}c@{}} $35.79^{+13.58}_{-11.97}$ \\ $26.5^{+8.98}_{-5.89}$ \end{tabular}   & \begin{tabular}[c]{@{}c@{}} $8.40^{+1.65}_{-1.04}$ \\ $11.98^{+2.57}_{-1.78}$ \end{tabular}   & 
  \begin{tabular}[c]{@{}c@{}} $8.93^{+0.93}_{-0.76}$ \\ $10.46^{+1.23}_{-0.80}$ \end{tabular} \\
  \hline

 $q$ &\begin{tabular}[c]{@{}c@{}}Standard  \\ PBH  \end{tabular}    & \begin{tabular}[c]{@{}c@{}} $0.86^{+0.12}_{-0.20}$ \\ $0.86^{+0.12}_{-0.17}$ \end{tabular}
  & \begin{tabular}[c]{@{}c@{}} $0.72^{+0.25}_{-0.27}$ \\ $0.77^{+0.21}_{-0.26}$ \end{tabular}   & \begin{tabular}[c]{@{}c@{}} $0.64^{+0.28}_{-0.28}$ \\ $0.91^{+0.08}_{-0.13}$ \end{tabular}       & \begin{tabular}[c]{@{}c@{}} $0.6^{+0.35}_{-0.24}$ \\ $0.47^{+0.32}_{-0.14}$ \end{tabular}   & \begin{tabular}[c]{@{}c@{}} $0.29^{+0.13}_{-0.07}$ \\ $0.63^{+0.30}_{-0.16}$ \end{tabular}  &
  \begin{tabular}[c]{@{}c@{}} $0.31^{+0.07}_{-0.05}$ \\ $0.46^{+0.11}_{-0.06}$ \end{tabular}  \\

  \hline
  $\chi_{\rm eff}$ &\begin{tabular}[c]{@{}c@{}}Standard  \\ PBH  \end{tabular}    & \begin{tabular}[c]{@{}c@{}} $-0.06^{+0.10}_{-0.13}$ \\ $0.00^{+0.00}_{-0.00}$ \end{tabular}
  & \begin{tabular}[c]{@{}c@{}} $-0.05^{+0.19}_{-0.22}$ \\ $0.00^{+0.00}_{-0.00}$ \end{tabular}   & \begin{tabular}[c]{@{}c@{}} $0.06^{+0.14}_{-0.05}$ \\ $0.00^{+0.00}_{-0.00}$ \end{tabular}       & \begin{tabular}[c]{@{}c@{}} $0.29^{+0.22}_{-0.27}$ \\ $0.00^{+0.01}_{-0.01}$ \end{tabular}   & \begin{tabular}[c]{@{}c@{}} $0.22^{+0.09}_{-0.11}$ \\ $0.00^{+0.00}_{-0.00}$ \end{tabular}  &   
  \begin{tabular}[c]{@{}c@{}} $0.20^{+0.06}_{-0.07}$ \\ $0.00^{+0.00}_{-0.00}$ \end{tabular} \\
  \hline
  \hline
   $z$ &\begin{tabular}[c]{@{}c@{}}Standard  \\ PBH  \end{tabular}    & \begin{tabular}[c]{@{}c@{}} $0.10^{+0.03}_{-0.04}$ \\ $0.09^{+0.03}_{-0.04}$ \end{tabular}
  & \begin{tabular}[c]{@{}c@{}} $0.26^{+0.09}_{-0.10}$ \\ $0.26^{+0.09}_{-0.11}$ \end{tabular}   & \begin{tabular}[c]{@{}c@{}} $0.07^{+0.02}_{-0.02}$ \\ $0.07^{+0.03}_{-0.03}$ \end{tabular}       & \begin{tabular}[c]{@{}c@{}} $0.29^{+0.07}_{-0.13}$ \\ $0.26^{+0.09}_{-0.13}$ \end{tabular}   & \begin{tabular}[c]{@{}c@{}} $0.17^{+0.03}_{-0.06}$ \\ $0.17^{+0.04}_{-0.07}$ \end{tabular}   &  
  \begin{tabular}[c]{@{}c@{}} $0.13^{+0.04}_{-0.05}$ \\ $0.12^{+0.05}_{-0.05}$ \end{tabular} \\
  \hline
  \hline

$\log{\cal B}$&\begin{tabular}[c]{@{}c@{}}Standard  \\ PBH  \end{tabular}    & \begin{tabular}[c]{@{}c@{}} $280.0\pm0.1$ \\ $279.0\pm0.2$ \end{tabular} 
  & \begin{tabular}[c]{@{}c@{}} $43.0 \pm 0.1$ \\ $42.3 \pm 0.1$ \end{tabular}   & \begin{tabular}[c]{@{}c@{}} $87.3 \pm 0.2$ \\ $88.7 \pm 0.2$ \end{tabular}        & \begin{tabular}[c]{@{}c@{}} $44.1 \pm 0.1$ \\ $40.1 \pm 0.1$ \end{tabular}   & \begin{tabular}[c]{@{}c@{}} $146.7 \pm 0.2$ \\ $144.8 \pm 0.2$ \end{tabular}  &
  \begin{tabular}[c]{@{}c@{}} $154.36 \pm 0.18$ \\ $149.15 \pm 0.16$ \end{tabular}  
 \\ \hline
\end{tabular}
\caption{Summary of the GW events analyzed with PBH-motivated priors ignoring accretion effects. Each event is representative of a class (low/moderate mass, (a)symmetrical, (non)spinning binaries). 
The last two columns refer to the analysis including only the dominant $l=m=2$ harmonic (GW190412*) and also higher harmonics (GW190412), as discussed in the text.
We show the median values of the binary parameters obtained with the standard (flat) priors on the (source-frame) masses and spins adopted by the LVC  and with the PBH-motivated priors neglecting accretion. The last row presents the (natural) log Bayes factors obtained in the two cases for each event. Errors represent the $90\%$ confidence intervals.
}
\label{tab:Log Bayes Factor}
\end{table}

\begin{figure*}
\includegraphics[width=0.9\textwidth]{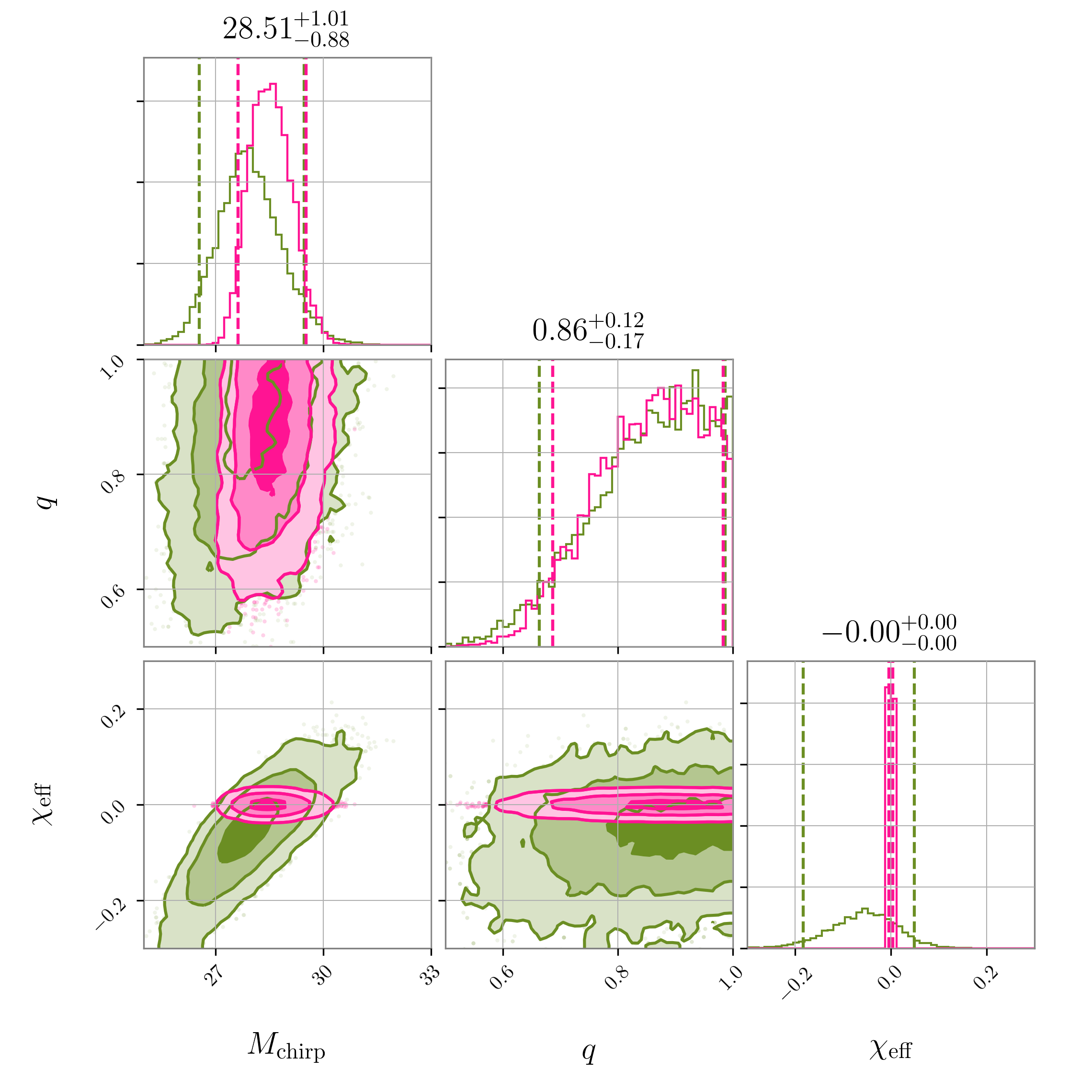}
 \caption{Corner plot for the relevant binary BH parameters of GW150914. The posterior distributions obtained with the PBH-motivated priors without accretion for the masses and spins correspond to the magenta contours, whereas the green curves correspond to the posterior obtained with standard priors used in the LVC analysis. The values quoted on the one-dimensional histogram correspond to the PBH-motivated priors.} 
 \label{fig:corner_GW150914}
 \end{figure*}

The results of the Bayesian inference for these systems performed by implementing the priors in Table~\ref{tab:Prior-tab} within the BILBY infrastructure are presented below.
Figure~\ref{fig:corner_GW150914} shows a corner plot with relevant parameters for GW150914, which is the prototypical example of a moderate-mass, symmetrical binary with a high signal-to-noise ratio~\cite{Abbott:2016blz}.
In the case of the standard priors, we recover the LVC results for the posterior distributions of the parameters (green curves). With the PBH-motivated priors without accretion (magenta curves), the effective spin is assumed to be essentially zero, so its posterior simply reflects the very narrow prior. In turn, this provides a slightly better measurement of the chirp mass, since the dimensionality of the waveform parameter space is effectively reduced. However, in the case of GW150914 the chirp mass is only mildly affected by the different choice of the priors (and anyway the new posterior is well within the $1\sigma$ contour of the standard one), whereas the posterior distribution of the mass ratio is almost unaffected.
This can be understood by noticing that the effective spin parameter of GW150914 as measured by the LVC is compatible with zero, so a narrow prior $\chi_\text{\tiny eff}\approx0$ is compatible with the measurement and does not affect the inference on the other parameters. Likewise, the chirp mass estimated for GW150914 is well within the best-fit lognormal distribution inferred for PBHs (see Fig.~\ref{fig:priors_no_acc}). Overall, for GW150914 the effect of the PBH-motivated priors without accretion is negligible. 
Although we do not explicitly show the inference plots for GW170818~\cite{LIGOScientific:2018mvr}, we find very similar results also for this system, which has the same qualitative features as GW150914.

 \begin{figure*}
\includegraphics[width=0.9\textwidth]{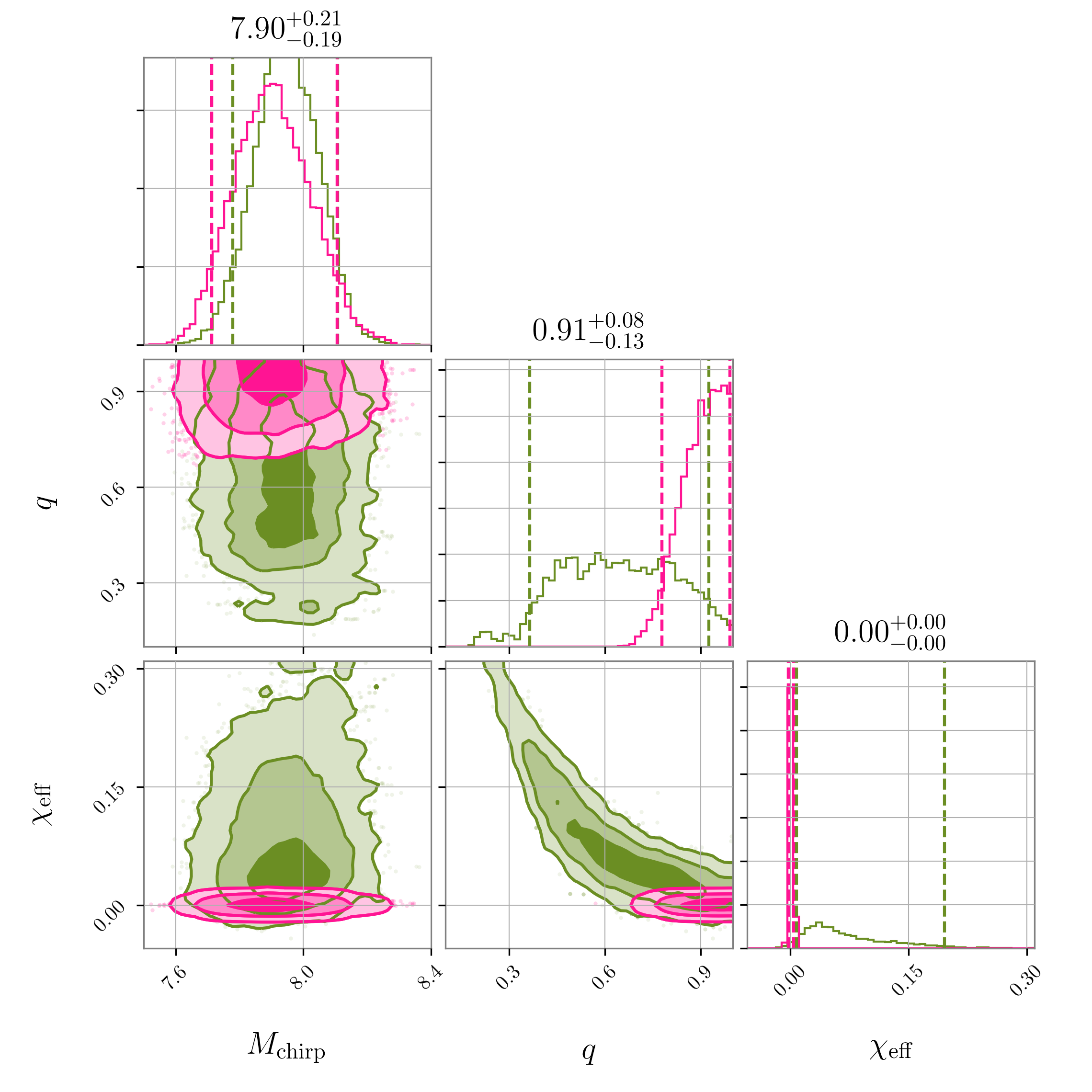}
 \caption{Same as in Fig.~\ref{fig:corner_GW150914} but for GW170608.}  
 \label{fig:corner_GW170608}
 \end{figure*}
 
The results of the analysis of GW170608~\cite{Abbott:2017gyy} are shown in Fig.~\ref{fig:corner_GW170608}. 
In this case we observe three effects: (i)~the standard posterior distribution of $\chi_\text{\tiny eff}$ is peaked slightly off zero, although $\chi_\text{\tiny eff}\approx0$ is still compatible with it and hence not necessarily in tension with the negligible-spin prior; (ii) besides being more narrow as in the case of GW150914, the mass distribution arising from the PBH prior is also peaked towards slightly smaller values compared to the standard prior case; (iii)~the combination of these effects conspires to impact on the posterior distribution of $q$ more significantly. In particular, the PBH priors yield a more accurate measurement which is skewed towards larger values (median $q=0.91^{+0.08}_{-0.13}$), although still compatible with the standard-prior case within the errors.

 \begin{figure*}
\includegraphics[width=0.9\textwidth]{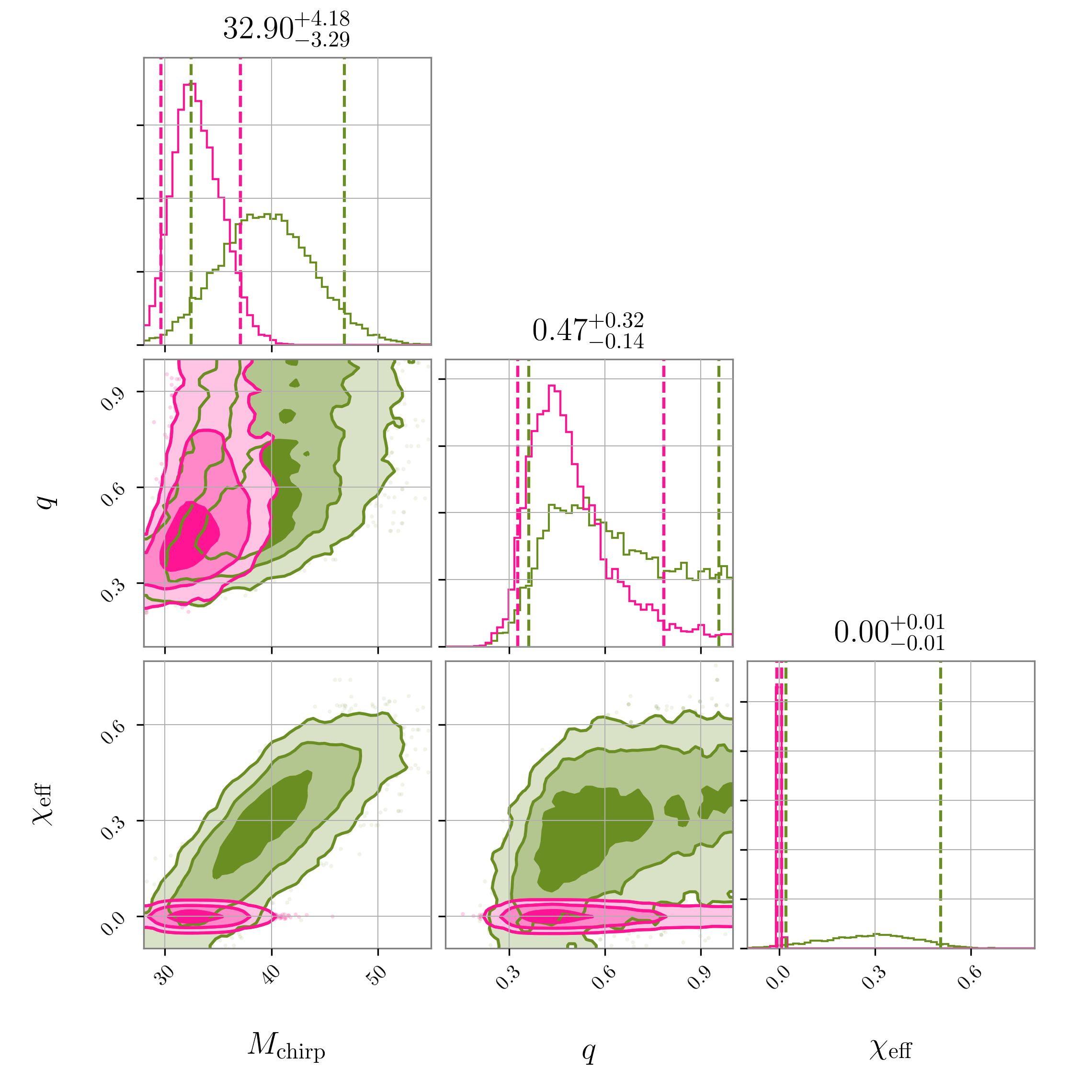}
 \caption{Same as in Fig.~\ref{fig:corner_GW150914} but for GW170729.}  
 \label{fig:corner_GW170729}
 \end{figure*}

In Fig.~\ref{fig:corner_GW170729} we show the results for GW170729~\cite{LIGOScientific:2018mvr}.\footnote{We note that in this case, using the standard LVC priors, we obtain posterior distributions which are slightly different from those reported by the LVC. However, uncertainties for this system are large and the inferred values significantly depend on the analysis, including the waveform model~\cite{Chatziioannou:2019dsz}.} In this moderate-mass case the standard analysis provides a moderate value for the median of $\chi_\text{\tiny eff}$ (although $\chi_\text{\tiny eff}=0$ is not excluded) and a mass ratio which is significantly smaller than unity (although with large uncertainties). Here, we observe the same qualitative effects as in the previous cases: namely the PBH-motivated priors yield a slightly lower chirp mass with smaller errors, and also affect the mass ratio which is correlated to $\chi_\text{\tiny eff}$.

\begin{figure*}
\includegraphics[width=0.49\textwidth]{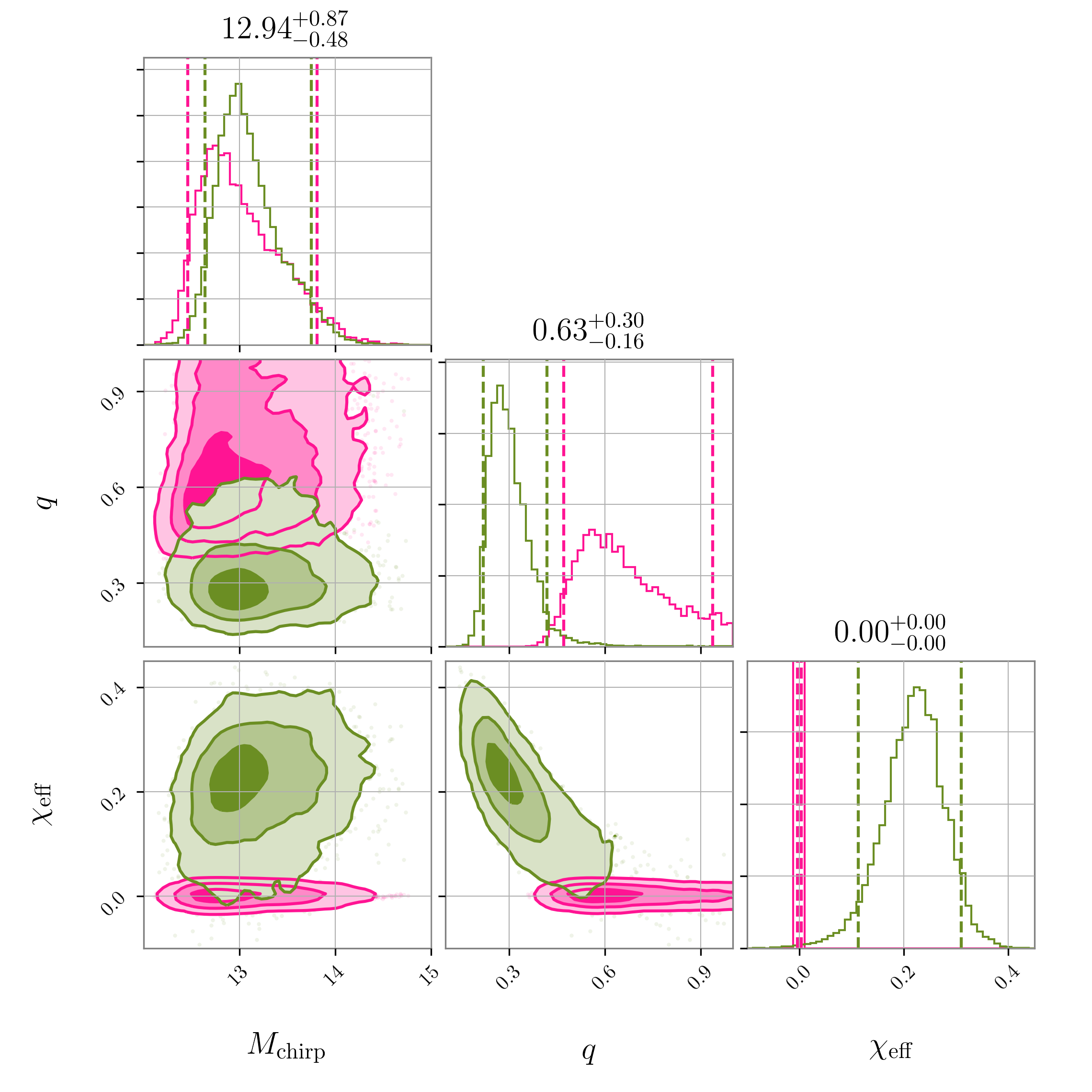}
\includegraphics[width=0.49\textwidth]{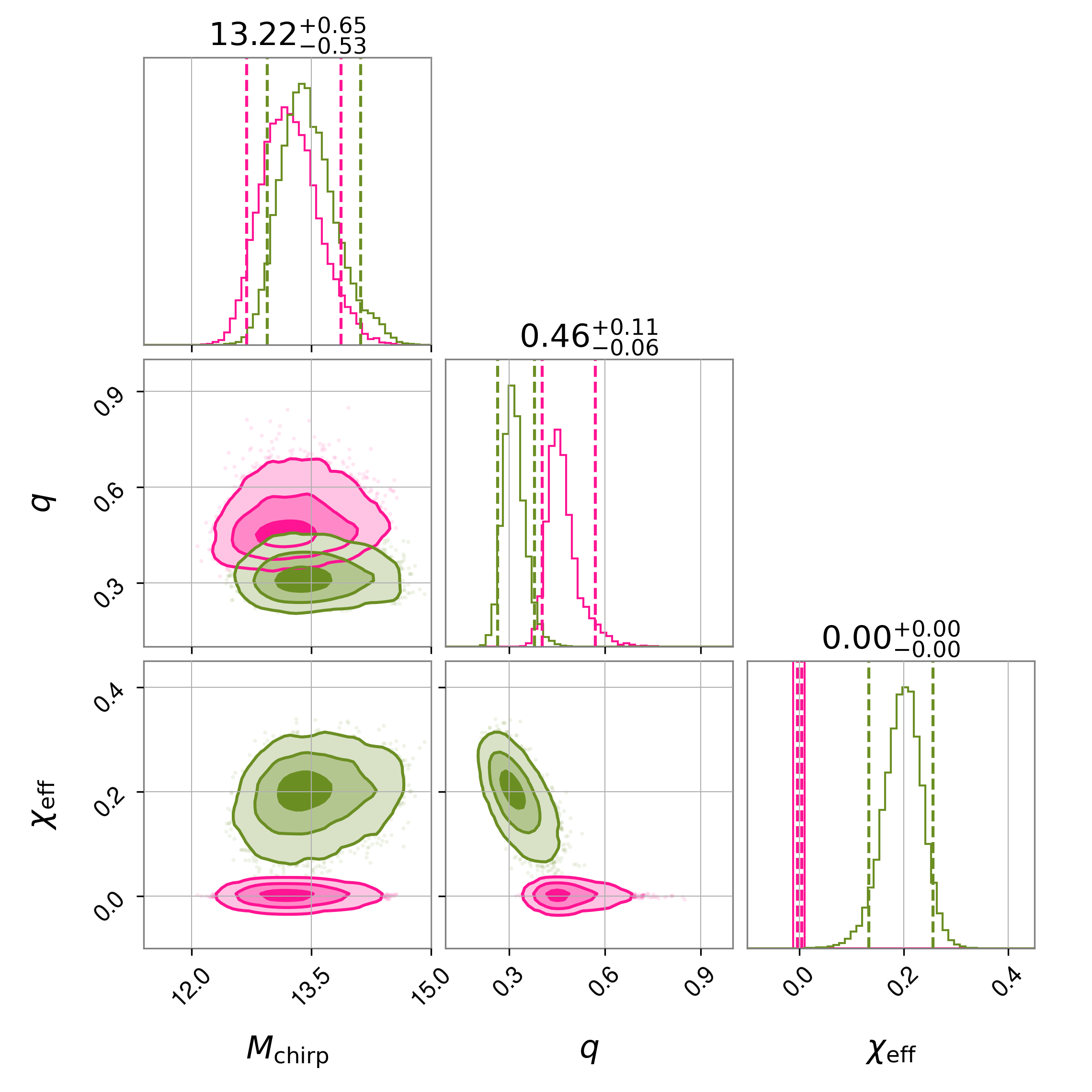}
 \caption{Same as in Fig.~\ref{fig:corner_GW150914} but for GW190412. The left (right) corner plot refers to the case in which higher harmonics in the waveform model are neglected (included) in the parameter estimation of this event. 
  }
  \label{fig:corner_GW190412}
   \end{figure*}

Finally, we now present the results for GW190412, the first binary BH event published from the O3 run~\cite{LIGOScientific:2020stg}. The main results are shown in Fig.~\ref{fig:corner_GW190412} for both the cases without (left panel) and with (right panel) the inclusion of higher harmonics in the waveform.
This event is particularly interesting for our study because it was confidently identified as a spinning ($\chi_\text{\tiny eff}=0.22^{+0.09}_{-0.11}$) and asymmetric ($q=0.29^{+0.13}_{-0.07}$) binary by the LVC analysis. However, these measurements rely on the standard (agnostic) choice for the mass and the spin priors, whereas a different prior assumption might change the qualitative nature of the inferred binary BH parameters~\cite{Mandel:2020lhv,Zevin:2020gxf}.

As compared to the other cases, GW190412 presents some peculiarities. Being a low-mass binary system, its chirp mass  inferred using the agnostic prior is also compatible with the PBH-motivated priors, so the two posteriors for $M_\text{\tiny chirp}$ as shown in Fig.~\ref{fig:corner_GW190412} are very similar, in both cases with (right) and without (left) the inclusion of higher harmonics.
However, for this event $\chi_\text{\tiny eff}=0$ is excluded roughly at $3.3\sigma$ confidence level if one adopts the standard flat priors on the spins. This is in tension with the PBH-motivated prior without accretion which imposes $\chi_\text{\tiny eff}\approx0$. Consequently, the measurement of the mass ratio is strongly affected in order to compensate for this tension, since $q$ and $\chi_\text{\tiny eff}$ are correlated in the waveform. Specifically, the distribution of $q$ broadens up and gains support at higher values. While with standard priors GW190412 can be confidently identified as a strongly asymmetric binary (in both cases with and without the inclusion of higher harmonics), the case of PBH-motivated priors is more involved. If one neglects the higher harmonics (left corner plot in Fig.~\ref{fig:corner_GW190412}), the PBH-motivated priors give a $q$ distribution which is also compatible (within less than $2\sigma$ confidence level) with a perfectly symmetric binary ($q=1$).
When the higher harmonics are included (right corner plot in Fig.~\ref{fig:corner_GW190412}), the distribution of $q$ is still shifted towards larger values relative to the agnostic-prior case, but it loses support near $q\approx 1$ with respect to the PBH-prior case with the dominant $l=m=2$ harmonic only. The reason can be understood by the fact that for GW190412 the subdominant $l=m=3$ mode was actually detected~\cite{LIGOScientific:2020stg}, which implies a certain degree of asymmetry for the system. The latter can arise mainly from $q\neq1$ and, to a less extent, from misaligned spin vectors. Since our waveform model for this analysis (IMRPhenomHM) assumes aligned spin vectors, the only asymmetry can come from the mass ratio of the system, which cannot be unity. Nonetheless, the qualitative result holds: imposing $\chi_{\rm eff}\approx0$ results in a larger inferred value of $q$ compared to the agnostic prior case.

We expect this to be a generic result that should apply to any binary for which the distribution of $\chi_\text{\tiny eff}$ (as inferred using standard priors) is incompatible with zero. If in addition the mass of the binary is large, there might be other effects due to different posterior distributions of $M_\text{\tiny chirp}$, unlike in the case of GW190412.
 
Although not shown in the corner plot, one of the parameters of the waveform model is the luminosity distance, from which the redshift of the source can be inferred. As shown in Table~\ref{tab:Log Bayes Factor}, the redshift is almost insensitive to the different choice of the priors.

Finally, one can compute the Bayes factors for the presence of a signal against noise, as explained in Sec.~\ref{sec:analysis}. These are presented in the last row of Table~\ref{tab:Log Bayes Factor} for the GW events considered above. Larger the value in this row, more is the preference of the data towards that prior model.
Interestingly enough, even neglecting accretion, for most of the cases the Bayes factors do not disfavor the hypothesis of the BH events being of primordial origin strongly. All events considered in our analysis --~with the exception of GW170608~-- show a weak statistical preference for the standard priors. For GW190412, the preference for the standard priors increases with the physically-motivated inclusion of higher harmonics in the waveform, with a difference in the (natural) log Bayes factor of $\sim 5$. This corresponds to the agnostic prior case being $\approx 150$ more likely than the (non-accreting) PBH hypothesis.
 
\subsection{With accretion}
As discussed in Sec.~\ref{sec:priors_accr}, an efficient phase of accretion during the cosmic history of PBHs affects the distribution of the individual masses and spins at detection~\cite{DeLuca:2020bjf,DeLuca:2020fpg}, and also modifies the best-fit values of a given mass function as obtained from the likelihood analysis of the observed events under the hypothesis that the latter are of primordial origin~\cite{DeLuca:2020qqa}.
Furthermore, in the case of accretion the spin prior distributions are correlated with those of the masses. Thus, while it is straightforward to implement the exact prior distributions for the masses, those of the spins are more involved. As explained in Sec.~\ref{sec:accr}, we adopted approximated spin distributions which account for the mass/spin correlation and are derived by fixing the chirp mass as inferred by a standard analysis. We expect this approximation to be accurate as long as $M_\text{\tiny chirp}$ does not change significantly with different prior choices, which can be checked a posteriori and is indeed our case.

The approximated spin prior distribution can be read off the panels in Fig.~\ref{fig:priors_acc_2}. For moderate-mass events (like GW150914, GW170818, and GW170729), the spin distribution peaks always near extremality and is very narrow (being these events above the mass-threshold for accretion), at least for small values of $z_{\co}$. In the case of moderate accretion ($z_{\co}=15$) the distribution of the spin is less extreme, but nonetheless centered around high values and without support at small spins. This implies that the prior distribution of $\chi_\text{\tiny eff}$ is broad for these systems (since the angles are unknown, see second row in Fig.~\ref{fig:priors_acc_2}) and can easily accommodate any measurement.
On the other hand, for light binaries (like GW170608) accretion is always inefficient and therefore, the spin distribution is similar to the non-accreting case, being sharply peaked at $\chi_{1,2}\approx0$. As show in Fig.~\ref{fig:priors_acc_2}, the case of GW190412 is in between these two regimes: in this case we expect that the spin distribution should depend strongly on $z_{\co}$: for $z_{\co}=15$ the spins are small as in the non-accreting case, whereas for the unrealistic case $z_{\co}=7$ the spins are moderately high with no support at $\chi_{1,2}=0$, see also  Fig.~\ref{fig:spin-acc-prior}.

\begin{figure*}[t]
\includegraphics[width=0.49\textwidth]{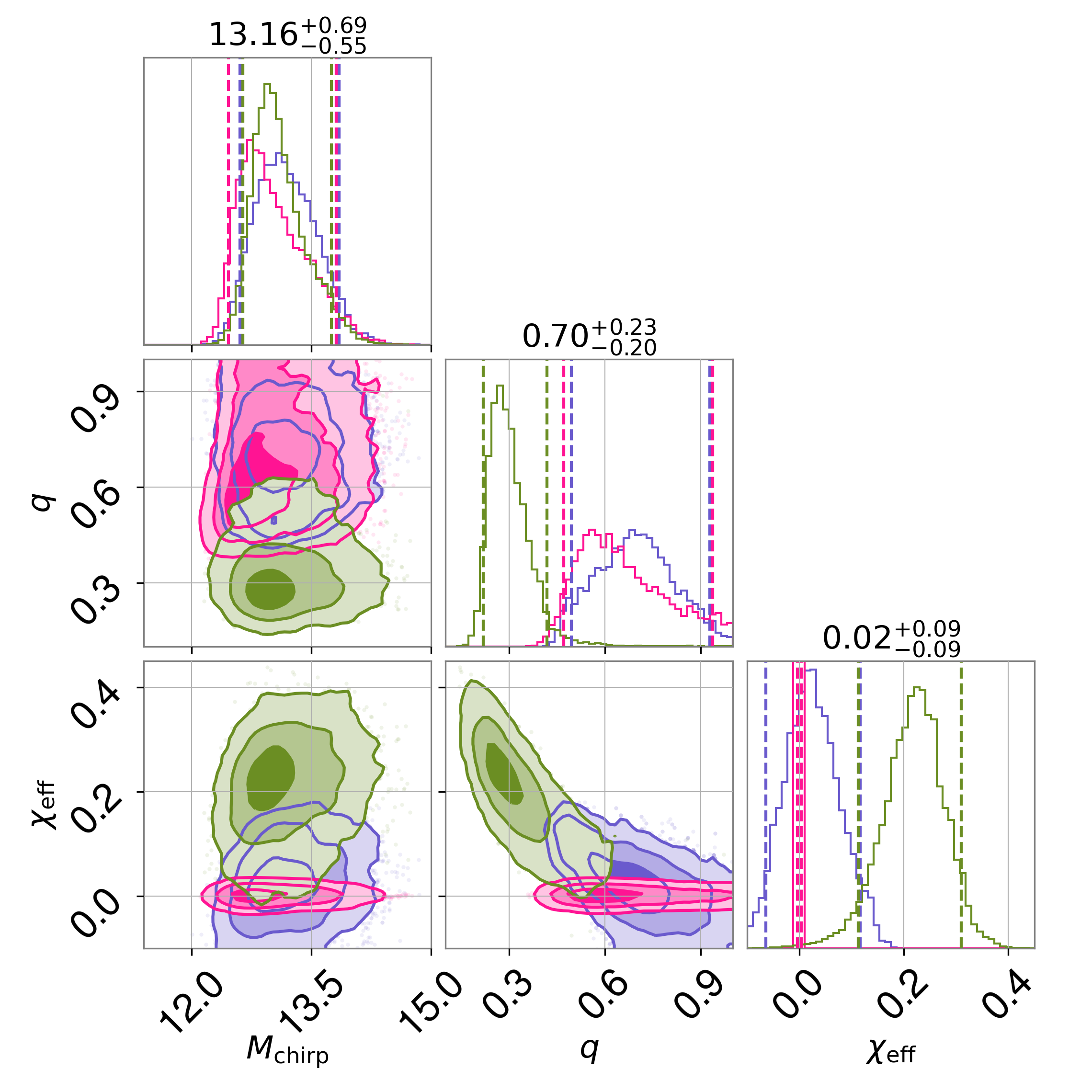} 
\includegraphics[width=0.49\textwidth]{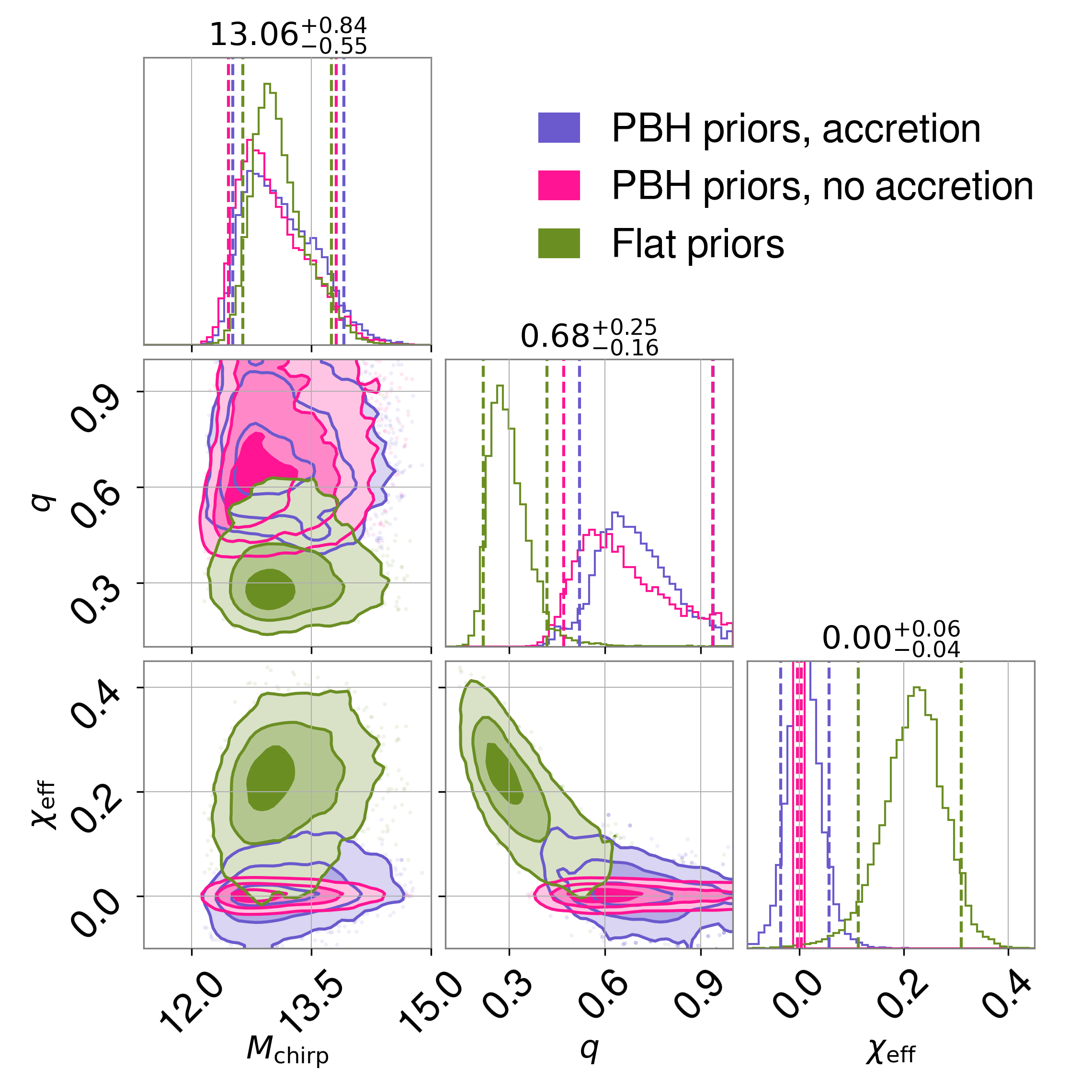}\\
\includegraphics[width=0.49\textwidth]{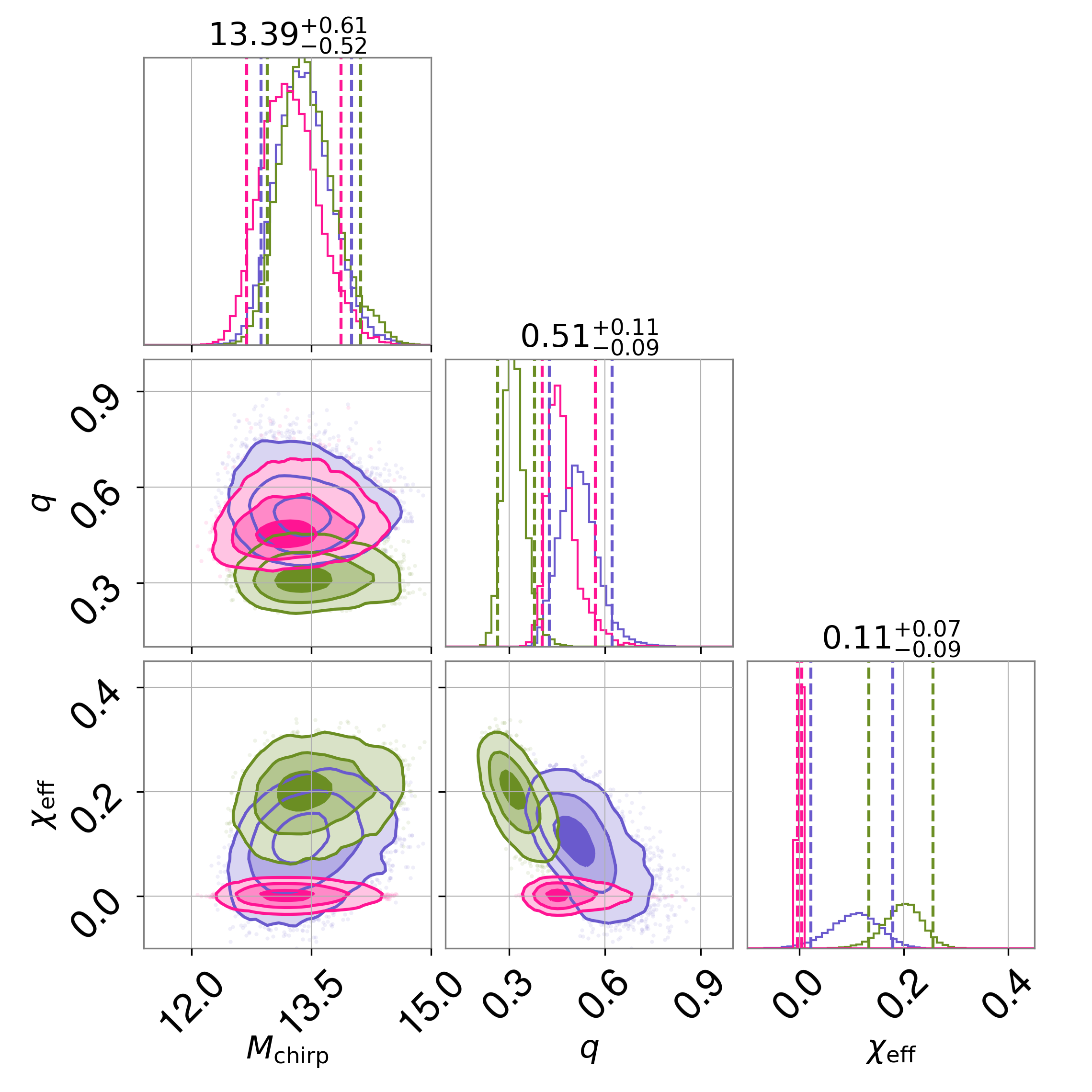} 
\includegraphics[width=0.49\textwidth]{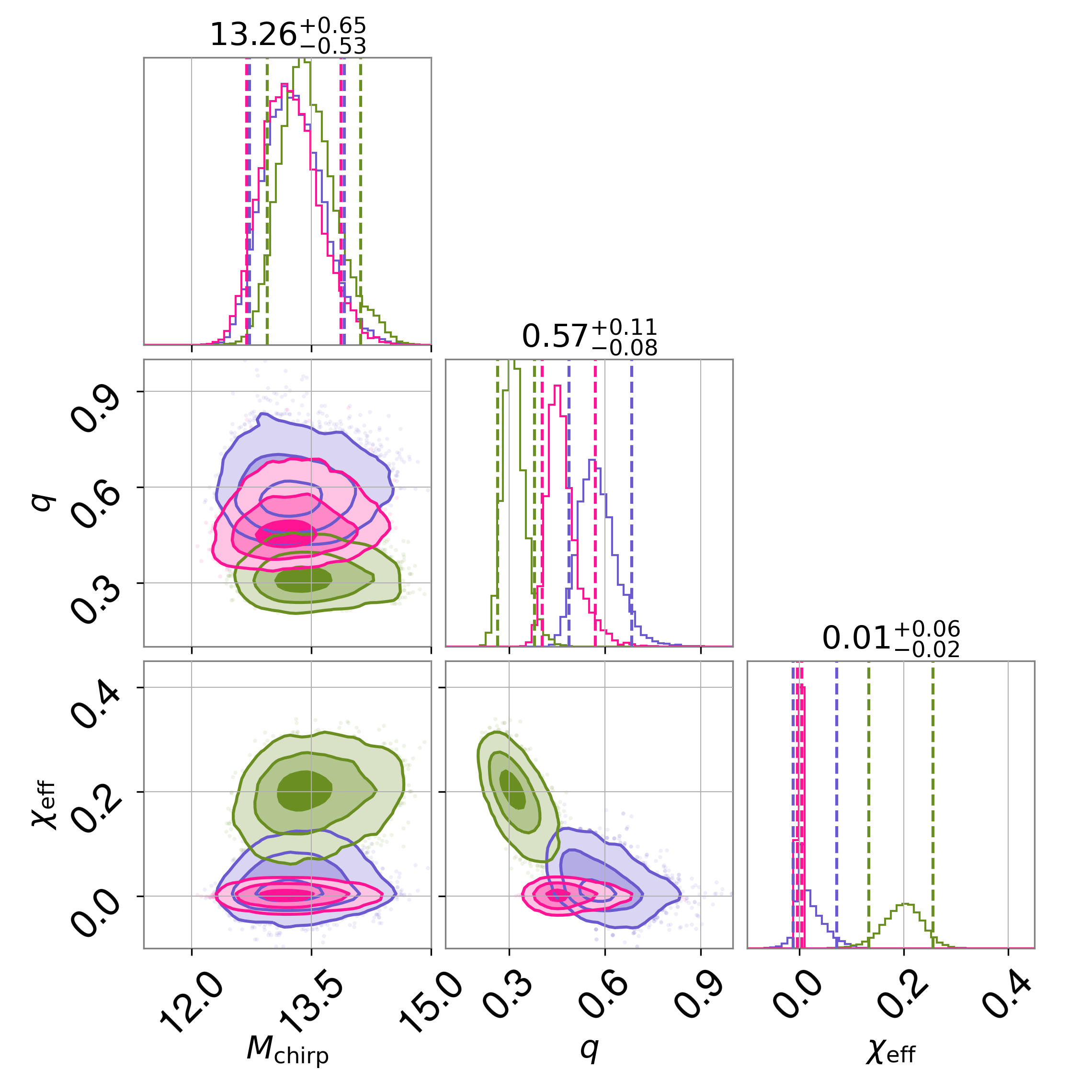}
 \caption{Corner plot for GW190412 in the case of PBH-motivated priors with accretion. The two top corner plots refer to the GW190412* analysis, i.e. neglecting higher harmonics in the waveform model, whereas the two bottom corner plots refer to the case in which higher harmonics are included for the parameter estimation. For both cases, the left (right) corner plot refers to an accretion cut-off $z_{\co}=10$ ($z_{\co}=15$).
 The purple curves correspond to the accreting PBH priors on masses and spins. For reference, the green curves correspond to standard (flat) priors, whereas the magenta curves correspond to the non-accreting PBH scenario (see Fig.~\ref{fig:corner_GW190412}). 
 The values on the top of the 1-d histograms correspond to the case where we use the full accretion priors (purple curves).} 
 \label{fig:corner_accretion}
 \end{figure*}

This preliminary analysis suggests that the effect of accretion can be understood by performing the parameter estimation for GW190412 with different values of $z_\co$ as representative examples. This is done in Fig.~\ref{fig:corner_accretion} for both the cases without (top corner plots) and with (bottom corner plots) the inclusion of the higher harmonics, respectively. For each case we consider the two most representative examples: accretion cut-offs $z_{\co}=10$ (left) and $z_\co=15$ (right).
The PBH-motivated priors including accretion correspond to the purple curves. For comparison, we also show the previous cases of standard agnostic priors (green curves) and of PBH-motivated priors without accretion (magenta curves).

Let us first focus on the case without higher harmonics (top corner plots of Fig.~\ref{fig:corner_accretion}).
For $z_{\co}=15$, accretion is small and the spin prior distributions are centered around zero (see right panels of Fig.~\ref{fig:priors_acc_2} and Fig.~\ref{fig:spin-acc-prior} and notice that the scale of the density function in Fig.~\ref{fig:priors_acc_2} is logarithmic). Thus, in this case we observe the same effect as in the non-accreting case. Namely, the posterior distribution of $\chi_\text{\tiny eff}$ is centered around zero (although is broader than in the non-accreting case) and --~as a consequence~-- the mass-ratio posterior moves to higher values closer to unity, with a distribution similar to the non-accreting PBH case.

On the other hand, already for $z_{\co}=10$ (top left corner plot of Fig.~\ref{fig:corner_accretion}) the effect of accretion is much stronger: in this case the individual spin distributions are typically high and yields a very broad distribution on $\chi_\text{\tiny eff}$ (see second row of Fig.~\ref{fig:priors_acc_2}). In this case, the constraining power of $\chi_\text{\tiny eff}$ is strongly reduced, since the prior can accommodate almost any value. Nonetheless, we observe that the posterior distribution of the mass ratio and of the effective spin are different from those obtained with the standard agnostic priors. This is due to the peculiar prior distribution of $q$: strongly accreting binaries tend to be symmetrical and therefore the priors on $q$ are skewed towards unity, being in tension with the measurement of $q$ inferred for GW190412 with standard priors. Thus, compared to the non-accreting PBH scenario (in which the posterior on $q$ is modified to compensate for a narrow prior on $\chi_\text{\tiny eff}$ which is in tension with the data), in the
%
strongly-accreting PBH scenario the opposite occurs: higher values of $q$ are favored by the priors and also $\chi_\text{\tiny eff}$ is modified to compensate for this difference. We expect this to be a generic feature for those binaries identified as highly asymmetrical using agnostic priors.

To check this expectation, we have repeated the parameter estimation of GW190412 using a {\it mixed}, unphysical choice of the priors: namely, we used the accreting PBH-motivated priors for the spins but standard (flat) priors for the masses. In this case the posterior distribution is very similar to the standard one, since the prior distributions are compatible with each other.

In all of the cases discussed, the posterior on $M_\text{\tiny chirp}$ is essentially unchanged, since this parameter is less strongly correlated with the others shown in the corner plot.
Furthermore, similarly to the non-accreting case, the redshift of the source inferred from the luminosity distance does non change significantly with the choice of PBH-motivated priors.

As shown in the bottom corner plots of Fig.~\ref{fig:corner_accretion}, the situation when including higher harmonics is similar, although with some important difference. We again find that the posterior distribution for $M_\text{\tiny chirp}$ and $\chi_\text{\tiny eff}$ with the accretion priors are very similar to the standard prior case. However, the posterior distribution of $q$ does not have support close to $q \sim 1$ when we use higher harmonics in the waveform. Furthermore, at variance with the dominant mode only analysis, the distribution of $z_\co=10, 15$ are visibly different when we include higher harmonics in the waveform.

Finally, let us look at the evidence for the various scenarios. Neglecting higher harmonics, the Bayes factors for GW190412 in the presence of accretion are $\log{\cal B}=(144.6\pm0.2,145.5\pm0.2)$ for $z_\co=(10,15)$, respectively. These are to be compared with those obtained with standard (flat) priors ($\log{\cal B}=146.7\pm0.2$) and with PBH-motivated priors in the absence of accretion ($\log{\cal B}=144.8\pm0.2$), see Table~\ref{tab:Log Bayes Factor}. Overall, neglecting higher harmonics the Bayes factors do not disfavor the primordial origin of GW190412 significantly, especially in the moderately-weak accretion case. 
On the other hand, when higher harmonics are included, the Bayes factors for GW190412 are $\log{\cal B}=(142.11\pm0.21,142.55\pm 0.19)$ for $z_\co=(10,15)$, to be compared with those given in Table~\ref{tab:Log Bayes Factor} for the higher-harmonics case  in the absence of accretion ($\log{\cal B}=149.15\pm0.16$) and with agnostic priors ($\log{\cal B}=154.36\pm0.18$). In this case we see that the agnostic priors are more favoured and the inclusion of higher harmonics increases this evidence.

Nonetheless, it is important to stress that the Bayes factors for the case of accretion have been calculated using the approximate priors. We caution the reader that this could introduce a systematic error in the evidence calculation depending on how the first approximation affects the whole prior volume 
(compared to the case of exact priors). However, quantifying this error would require a separate work and is beyond our scope here.
Furthermore, for the higher-harmonics analysis we assumed spins orthogonal to the orbital plane and, as clear from the above discussion, the evidence calculation is quite sensitive to various modelling assumptions, so the quantitative result might slightly change when including spin misalignment and orbital precession.

\section{Conclusions}
\label{sec:conclusion}
We have explored how assuming that (at least some of) the binary BH coalescences detected so far by the LVC have a primordial (rather than astrophysical) origin affects the binary parameters inferred through the GW data.
PBHs are likely to be formed with a negligible spin and might possibly acquire some angular momentum only if an efficient phase of accretion at the (super)Eddington rate occurs during their cosmic history~\cite{DeLuca:2020bjf,DeLuca:2020qqa}. These properties change the prior distributions of BH masses and spins relative to the agnostic (flat) distributions typically adopted.
We performed a full Bayesian parameter estimation using PBH-motivated priors for some representative transient GW signals. In overall agreement with previous work~\cite{Vitale:2017cfs,Mandel:2020lhv,Zevin:2020gxf}, our results show that the choice of the priors can affect the measurements (and possibly the physical interpretation) of the binary BH parameters significantly for those events in which the standard LVC measurements are in tension with the new prior distribution, since the waveform parameters tend to ``readjust" in order to alleviate the tension with the new priors.

Our analysis allows to draw some general conclusions:
 \begin{itemize}
    \item The chirp mass of the binary and the source redshift are almost insensitive to the choice of a PBH-motivated priors, whereas the posterior distribution of the mass ratio and of the effective spin can be affected significantly for some binary BH systems.
    \item A prior with negligible-spin support (as suggested by the PBH scenario without accretion, or by a putative astrophysical scenario in which the supernova remnant is slowly spinning~\cite{Fuller:2019sxi}, or also if quantum effects at BH formation are relevant~\cite{Bianchi:2018ula}) can strongly affect the overall parameter estimation of those binaries identified as spinning with a standard analysis that uses flat spin priors. In particular, at least for the cases we studied, the posterior distribution of the mass ratio gains more support close to unity relative to the standard case. Thus, binaries identified as (highly) spinning and (highly) asymmetrical by using standard agnostic priors might actually be non-spinning and symmetrical. Interestingly enough, for most of the events considered in this work the Bayes factor do not strongly favor one hypothesis against the other.
    \item A strong phase of accretion in the PBH history yields a bimodal prior distribution of the spin magnitudes, which is also mass- (and redshift-) dependent. Nonetheless, the distribution of $\chi_\text{\tiny eff}$ is typically broad, owing to the broadness of the distribution of the individual masses (which propagates into $\chi_\text{\tiny eff}$, see Eq.~\eqref{chieff}) and --~most importantly~-- because the spin orientations are unknown. Even if both the masses and the spin magnitudes were known with infinite precision, the uncertainty on the spin orientations makes the prior distribution of $\chi_\text{\tiny eff}$ broad, jeopardizing its constraining power in the case of strong accretion.
    \item Nonetheless, in the accreting case BH binaries tend to be symmetrical, so the prior on the mass ratio $q$ is skewed towards larger values. This might strongly affect the inference on $q$ (and on $\chi_\text{\tiny eff}$, which is correlated to it) for those binaries identified as asymmetrical using agnostic priors, which is the case of GW190412.
    \item In particular, one of the main results of our analysis is that, if GW190412 is assumed to be of primordial origin, its mass ratio inferred from the data is larger than in the standard agnostic-prior case, the extend to which depends on the waveform model. In particular, neglecting higher harmonics in the waveform even $q=1$ is compatible within $2\sigma$. When higher harmonics are included (but assuming aligned spins) the posterior of $q$ loses support near $q\approx1$, since the detection of a $l=m=3$ mode for GW190412 requires some asymmetry in the system. Finally, also the evidence is slightly affected by the inclusion of the higher harmonics.
    \item The latter discussion also shows relevance of an accurate waveform modelling for GW190412, since any modelling systematics can in principle affect the evidence. Indeed, we highlight that modelling the waveform accurately is essential to perform model selection. Whilst the log Bayes factor is comparable for the PBH prior and standard prior for GW190412 including only the dominant harmonic, inclusion of higher harmonics breaks this degeneracy. Adding higher harmonics to model selection between the agnostic and PBH scenario for GW190412 is therefore crucial.
\end{itemize}

Many qualitative aspects of our analysis are valid beyond the PBH scenario. In particular, the above conclusions for the non-accreting PBH case (in which the priors implement the information that the PBH spins are essentially negligible) are also qualitatively valid for other scenarios in which the binary BHs are not necessarily of primordial origin but their spin is nonetheless small. In this case, we predict that the mass ratio of binaries identified as spinning using standard (flat) spin priors should be strongly affected and tend more towards unit to compensate for the absence of an effective spin term in the waveform model.

We also  note that even if in the PBH scenario the physical parameters of some binaries are different from the standard ones, they do not affect a $\chi^2$-analysis of the PBH mass function significantly, so the analysis of Ref.~\cite{DeLuca:2020qqa} is not affected by our results.

Although we considered only a subset of the O1-O2-O3 LVC events, our analysis allows us to identify the key factors which are responsible for the impact of the priors also in other events. For example, the recent GW190814~\cite{Abbott:2020khf} is a low-mass (with lighter companion in the low-mass gap and therefore prone to a PBH interpretation~\cite{Clesse:2020ghq}), asymmetrical, non-spinning binary. Therefore, we expect that in this case the PBH-motivated priors should not affect the binary parameters significantly when accretion is negligible, whereas the inferred mass ratio can be compatible with unity if accretion is strong. 
On the other hand, the LVC has recently released GW190521, the first binary BH with at least a component's mass in the mass gap predicted by the pair-instability supernova theory~\cite{Abbott:2020tfl,Abbott:2020mjq} (see also Ref.~\cite{Graham:2020gwr} for a candidate electromagnetic counterpart of this merger). GW190521 is a nearly symmetrical binary ($q=0.79^{+0.19}_{-0.29}$) likely with non-negligible spins ($\chi_1=0.69^{+0.27}_{-0.62}$, $\chi_2=0.73^{+0.24}_{-0.64}$), although the non-spinning scenario cannot be excluded due to the large measurement errors~\cite{Abbott:2020mjq}. This event is unique in many ways, in particular the spin vectors are oriented such that the effective spin is almost zero, and data favor the possibility that the binary is precessing. Interestingly, in the absence of accretion, the possibility that GW190521 is of primordial origin is excluded by other constraints, whereas such an interpretation is viable in the PBH accreting scenario~\cite{1814836}, which is also compatible with the above values of $q$ and $\chi_{1,2}$ inferred with agnostic priors.
A detailed analysis of these very interesting events, as well as all others in the current~\cite{Abbott:2020niy} and upcoming O3 catalogues, is left for future work.

An important extension of our work is to repeat the parameter estimation of GW190412 using a precessing waveform model which includes higher harmonics and check whether the inclusion of spin misalignment could modify the posteriors and the evidence analysis.

Finally, it would be interesting to compare the odds of the PBH hypothesis with those of different astrophysical scenarios, e.g. hierarchical mergers~\cite{Gerosa:2017kvu,Gerosa:2020bjb}, within the Bayesian framework used here.

\acknowledgments
\noindent 
We are indebted to Michael Zevin who pointed out to us the importance of higher harmonics for the evidence analysis of GW190412.
We thank the developers of BILBY code infrastructure, especially Greg Ashton, Colm Talbot, Moritz Hübner and Sylvia Biscoveanu for technical discussions on BILBY. 
Computations were performed at Sapienza University of Rome on the Vera cluster and at the University of Geneva on the Baobab cluster. We are also indebted to Francesco Pannarale for interesting discussion, and to Luca Graziani for computational support on the Vera Cluster.
V.DL., G.F. and A.R. are supported by the Swiss National Science Foundation 
(SNSF), project {\sl The Non-Gaussian Universe and Cosmological Symmetries}, project number: 200020-178787.
P.P. acknowledges financial support provided under the European Union's H2020 ERC, Starting
Grant agreement no.~DarkGRA--757480, under the MIUR PRIN and FARE programmes (GW-NEXT, CUP:~B84I20000100001), and 
support from the Amaldi Research Center funded by the MIUR program `Dipartimento di Eccellenza" (CUP:~B81I18001170001).






\bibliographystyle{JHEP}
\bibliography{Refs_PBHs}

\end{document}